\documentclass[aps,pra,reprint,showpacs,showkeys,groupedaddress]{revtex4-1}

\usepackage[dvips]{graphicx}
\usepackage{epsfig}
\usepackage{amssymb}
\usepackage{amsmath}
\usepackage{color}

\begin{document}
\title{Single-color two-photon spectroscopy of Rydberg states in electric fields}
\date{\today}
\author{T. E. Wall}
\author{D. B. Cassidy}
\author{S. D. Hogan}
\affiliation{Department of Physics and Astronomy, University College London, Gower Street, London WC1E 6BT, U.K.}
\pacs{32.70.Jz,32.80 Ee,32.80.Rm}

\begin{abstract}
Rydberg states of atomic helium with principal quantum numbers ranging from $n=20$ to $n=100$ have been prepared by non-resonance-enhanced single-color two-photon excitation from the metastable $2\,^3\text{S}_{1}$ state. Photoexcitation was carried out using linearly and circularly polarized pulsed laser radiation. In the case of excitation with circularly polarized radiation, Rydberg states with azimuthal quantum number $|m_{\ell}|=2$ were prepared in zero electric field, and in homogeneous electric fields oriented parallel to the propagation axis of the laser radiation. In sufficiently strong electric fields, individual Rydberg-Stark states were resolved spectroscopically, highlighting the suitability of non-resonance-enhanced multiphoton excitation schemes for the preparation of long-lived high-$|m_{\ell}|$ hydrogenic Rydberg states for deceleration and trapping experiments. Applications of similar schemes for Doppler-free excitation of positronium atoms to Rydberg states are also discussed.
\end{abstract}

\maketitle

\section{\label{sec:intro}Introduction}
Among their many often extreme properties, Rydberg states of atoms and molecules can possess long radiative lifetimes. In the absence of external perturbations low-angular-momentum hydrogenic Rydberg states with principal quantum number $n>30$, and small quantum defects, typically exhibit lifetimes exceeding $1~\mu$s which scale with $n^3$. High-angular-momentum `circular' states with similar values of $n$ have lifetimes greater than $5~$ms which scale with $n^5$~\cite{gallagher94a}. Because of these long lifetimes, optical and microwave transitions involving these states exhibit narrow natural linewidths. They are therefore very well suited for precision spectroscopic measurements, including for example, the accurate determination of molecular ionization and dissociation limits~\cite{liu09a}, and studies of the role of nuclear spins in photoionization~\cite{paul09a,sassmannshausen13a}. These long lifetimes are also of importance for applications of Rydberg states in quantum information processing~\cite{gleyzes06a,saffman10a}, and in experiments with antihydrogen and positronium atoms directed toward (i) spectroscopic investigations of matter--antimatter asymmetries~\cite{bluhm99a}, and (ii) measurements of the acceleration of particles composed of antimatter in the gravitational field of the Earth~\cite{mills02a,kellerbauer08a,cassidy14a}. 

In each of these areas, the long radiative lifetimes of the Rydberg states alone are generally not sufficient for measurements at the highest resolution. It is also necessary to ensure that the period of time during which the samples are interrogated, and the interaction time with the desired external fields approach, or exceed, these lifetimes, and that motional effects such as Doppler broadening are minimized. Typically these requirements can be fulfilled simultaneously through the preparation of slowly moving samples with low translational temperatures. However, there are cases where high resolution excitation to Rydberg states directly from a ground, or other low-lying electronic, state is necessary before subsequent deceleration or trapping, because of the challenges associated with the preparation of sufficiently cold ground state samples using current techniques. These cases include, for example, precision spectroscopic studies of the ionization and dissociation limits of H$_2$ and its isotopomers~\cite{liu09a}, and the preparation of long-lived excited states of positronium~\cite{cassidy12a}. In these cases, it is foreseen to minimize effects of Doppler broadening in the photoexcitation process by using counter-propagating laser beams to drive non-resonance-enhanced single-color two-photon transitions to the Rydberg states~\cite{sprecher11a,hogan13a}. The spectral resolution of the resulting Doppler-free excitation~\cite{Haensch75a} will then be limited only by the bandwidth of the laser radiation used to drive the transition, and the period of time with which it interacts with the sample.

The photoexcitation of Rydberg states using non-resonance-enhanced single-color two-photon schemes in the presence of homogeneous electric fields offers the opportunity of exploiting the large electric dipole moments associated with the resulting states for efficient acceleration, deceleration and trapping using inhomogeneous electric fields~\cite{yamakita04a,vliegen04a,hogan08a}. The magnitude of these dipole moments scale with $n^2$, and at $n=30$ can exceed 3000~D. In addition to the importance of the results presented here for Doppler-free Rydberg state photoexcitation, when driven using circularly polarized laser radiation such multiphoton excitation schemes represent a general approach to the preparation of long-lived non-core-penetrating Rydberg states of small molecules. With their small quantum defects, and long fluorescence lifetimes, these states are ideally suited for deceleration and electric trapping experiments~\cite{hogan09a,seiler11a}, directed toward studies of excited state decay processes, and low energy scattering with other gas-phase targets, or surfaces~\cite{softley04a}.

Non-resonance-enhanced single-color two-photon excitation schemes have been employed previously in high resolution spectroscopic studies of atomic Rydberg states~\cite{stoicheff79a,penent88a} and in studies of (2+1) resonance-enhanced multiphoton ionization~\cite{genevriez14a}. We report here a detailed investigation of the role of ac Stark energy shifts in single-color two-photon excitation in beams of metastable helium (He) atoms under experimental conditions which are similar to those expected in future experiments with positronium (Ps), and report the efficient excitation of selected Rydberg-Stark states.

In the following, a description of the experimental apparatus used in the work reported here is first provided. The numerical methods employed in the calculation of the spectral profiles associated with non-resonance-enhanced two-photon transitions from the 2\,$^3\text{S}_1$ state in He to high Rydberg states is then outlined, before the results of the experiments carried out (i) in the absence of externally applied electric fields, and (ii) in the presence of homogeneous electric fields oriented parallel to the direction of propagation of the laser radiation, are presented. Finally, comparisons are drawn between the single-color two-photon spectra reported here and those expected in the case of Doppler-free two-photon excitation of positronium using laser radiation with similar temporal and spectral properties. 

\section{\label{expSet}Experiment}
\begin{figure}[b]
\centerline{\includegraphics[width=1.0 \linewidth]{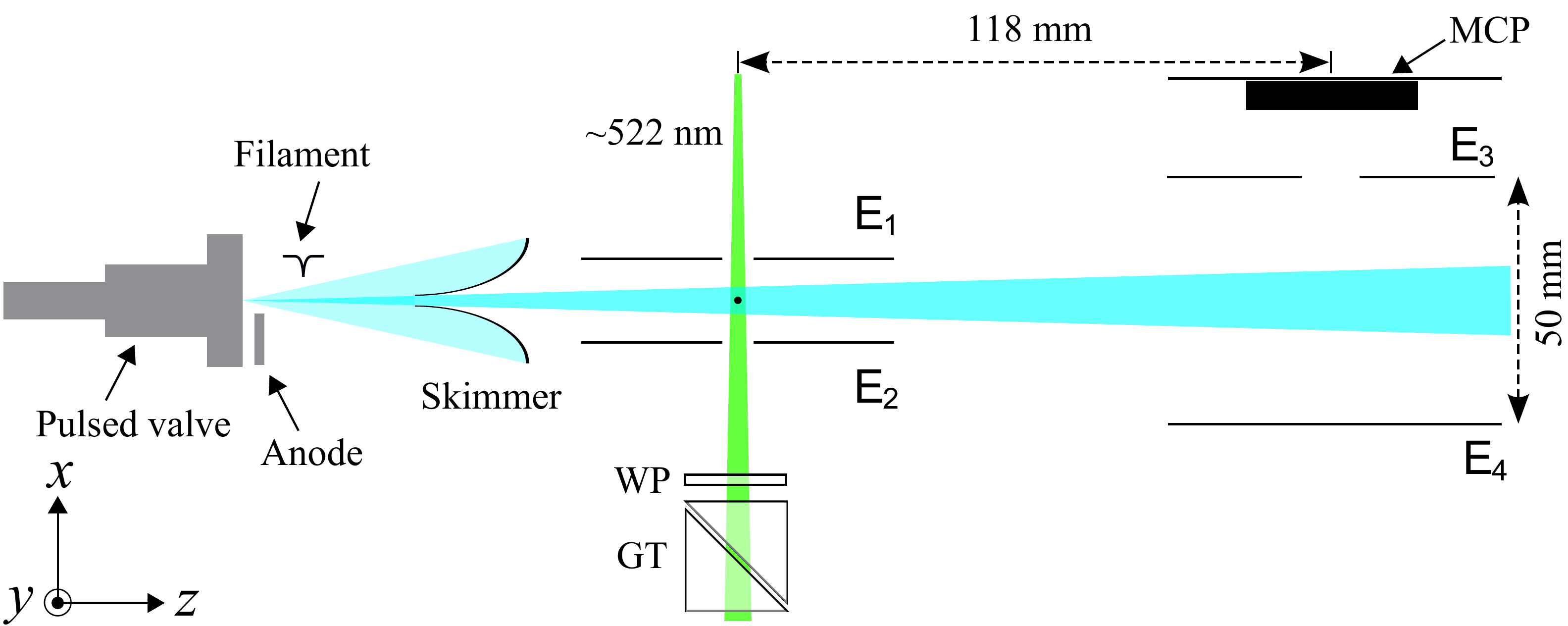}}
\caption{(Color online) Schematic diagram of the experimental setup (not to scale). The polarization of the laser beam is set by a Glan-Taylor polarizer (GT) and then controlled with a rotatable quarter-wave plate (WP). The coordinate system is defined such that the origin is located at the position where the laser beam axis intersects the atomic beam axis (shown as a black point)}
\label{fig:expSetup}
\end{figure}
A beam of metastable He atoms is prepared in an electric discharge within a supersonic expansion, as described in \cite{Halfmann00}. Helium is injected into the source vacuum chamber of the experimental apparatus through a pulsed valve (Series 9 General Valve). At the exit of the valve the metastable 2\,$^3$S$_1$ state is populated in an electric discharge generated between a metal anode located $\approx 1~$mm downstream from the nozzle to which an electric potential of $+240~$V is applied, and the grounded base plate of the valve. The discharge is seeded with electrons from a heated tungsten filament located $\approx 15~$mm further downstream. After expanding and cooling the metastable He beam travels with a mean forward speed of $v_z\simeq 1950~$ms$^{-1}$. Atoms in the shorter-lived 2\,$^1$S$_0$ metastable state also populated in the discharge, although present in the beam, do not play a noticeable role in the studies of the triplet states described here.

Having passed through a skimmer with a $2~$mm-diameter aperture into a second vacuum chamber, the atoms pass between  two parallel planar electrodes (labelled E$_1$ and E$_2$ in Fig.~\ref{fig:expSetup}) which each have a length of $100~$mm in the direction of propagation of the atomic beam and are separated by $18~$mm. Each electrode contains a $3~$mm aperture in the center. A single laser beam ($\sim10$~mJ per pulse at $\approx522~$nm with a pulse length of 6~ns and a bandwidth of 5~GHz, produced by a Nd:YAG-pumped pulsed dye laser) passes through these apertures. The laser beam converges as it propagates through the chamber, reaching a focus $\sim20~$mm beyond the atomic beam axis. Where the laser intersects the atomic beam axis it has a Gaussian transverse intensity profile, with a standard deviation $\sigma_y\simeq\sigma_z\simeq70~\mu$m (see Fig.~\ref{fig:expSetup} for the definition of the coordinate system), resulting in peak intensities on the axis of the atomic beam of $\sim5\times 10^{13}~$Wm$^{-2}$. By applying a potential difference between E$_1$ and E$_2$, uniform electric fields of up to $F=155~$Vcm$^{-1}$ can be produced in the excitation region, allowing the selective preparation of individual Rydberg-Stark states.
 
\begin{figure}[t]
\centerline{\includegraphics[width=1.0 \linewidth]{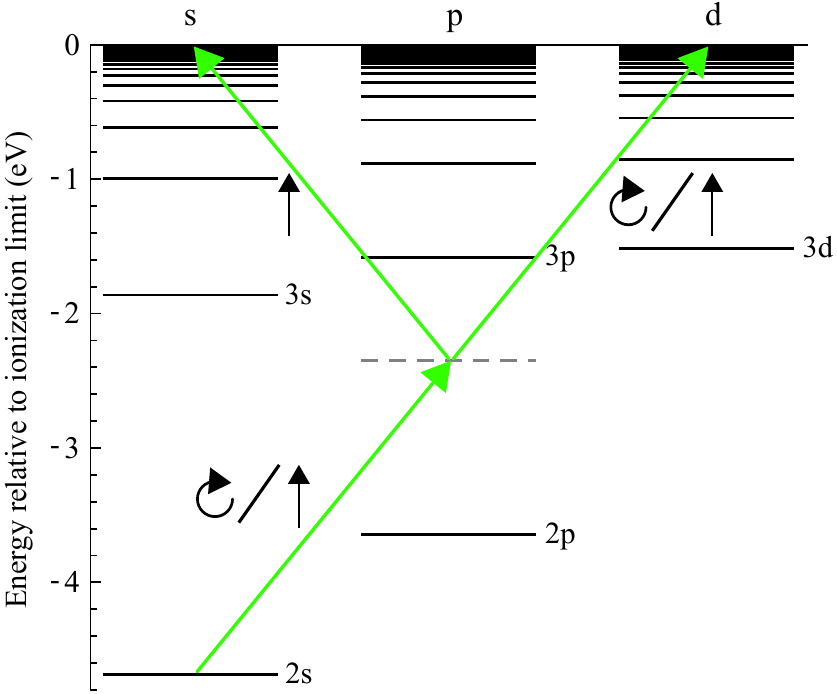}}
\caption{(Color online) Energy level diagram showing possible single-color two-photon excitation pathways from the 1s2s\,$^3\text{S}_1$ state to high Rydberg states of He using circularly ($\circlearrowright$) and linearly ($\uparrow$) polarized laser radiation.}
\label{fig:HeLevels}
\end{figure}

Single-color two-photon transitions driven from the 2\,$^3\text{S}_1$ state in He to Rydberg states with $n\geq 20$ converging on the $1\text{s}\,^{2}\text{S}_{1/2}$ ionization limit require radiation with wavelengths in the range from 520~nm to 524~nm (Fig.~\ref{fig:HeLevels}). With the intermediate state far from resonance with any bound state, these transitions are not resonance-enhanced.
In the experiment the laser polarization is set using a Glan-Taylor polarizer and then controlled using a $\lambda/4$-plate, permitting the production of circularly polarized radiation with a purity of $>90\%$. In the excitation region in the apparatus the electrode planes are oriented parallel to the axis of the atomic beam and perpendicular to the direction of propagation of the laser beam, such that the circularly polarized radiation drives transitions for which $\Delta$$M_J=+1$ or $\Delta$$M_J=-1$, where $M_J$ is the azimuthal quantum number associated with the total angular momentum $\vec{J}$, of the atom. In the case of the triplet states of He treated here only the interaction of the optically active electron, with total angular momentum $\vec{j}$, and azimuthal quantum number $m_j$, initially in the 2s orbital need be considered. Therefore excitation with circularly polarized radiation results in transitions for which $\Delta$$m_j=+1$ or $\Delta$$m_j=-1$. Since individual spin-orbit components are not resolved in the experiments described, in general the azimuthal quantum number, $m_{\ell}$, associated with the orbital angular momentum of the single optically active electron is referred to when discussing the effects of laser polarization in the observed spectra. With circularly polarized radiation driving single-photon transitions for which $\Delta$$m_{\ell}=+1$ or $\Delta$$m_{\ell}=-1$, two-photon transitions from the 1s2s\,$^3\text{S}_1$ state can result only in the excitation of states with $|m_{\ell}|=2$. However, linearly polarized radiation can drive transitions for which $\Delta$$m_{\ell}=\pm$1. Thus, in the configuration presented, linearly polarized radiation can drive two-photon transitions to states with $m_{\ell}=0$ and with $|m_{\ell}|=2$.

Following photoexcitation He Rydberg atoms are detected by pulsed electric field ionization 118~mm downstream from the excitation region between electrodes E$_3$ and E$_4$. The resulting ions are then collected on a micro-channel plate (MCP) detector. With E$_3$ and E$_4$ separated by 50~mm, the pulsed potential of $+4.0~$kV applied to E$_4$ for electric field ionization results in a field on the atomic beam axis of $\sim800$~Vcm$^{-1}$. This field is sufficient for complete diabatic ionization of atoms in states with principal quantum number $n\geq 35$ \cite{gallagher94a}. The signal from the He Rydberg atoms is recorded by integrating the gated output from the MCP, with each data-point the average of 30 pulses. The experiment is run at a repetition rate of 10~Hz to match that of the Nd:YAG laser pumping the dye laser.

\section{AC Stark shifts - calculation and lineshapes}\label{acStarkSec}

The strong laser fields required to drive the non-resonance enhanced two-photon transitions reported here, give rise to ac Stark shifts (light shifts) of the associated energy levels. These energy shifts lead to broadenings and asymmetries of the corresponding spectral features. The magnitudes of these effects depend upon the frequency-dependent dynamic polarizability, $\alpha(\omega)$, of each of the states between which the two-photon transitions are driven. In a laser field with an intensity $I_{\text{laser}}$, and an angular frequency $\omega_{\text{laser}}$, the ac Stark energy shift, $E_{\text{ac}}$, experienced by an atom in an electronic state with principal quantum number $n$, total angular momentum $\vec{J}=\vec{L}+\vec{S}$ (where $\vec{L}$ is the total electronic orbital angular momentum, and $\vec{S}$ is the total electron spin) and azimuthal quantum number $M_J$, is~\cite{Deissler08a}
\begin{eqnarray}
E_{\mathrm{ac}} &=& \frac{1}{2}\alpha_{|nJM_J\rangle}(\omega)\langle F_{\text{laser}}^2\rangle
\end{eqnarray}
where $\langle F_{\text{laser}}^2\rangle=I_{\text{laser}}/(\epsilon_0 c)$ is the root-mean-square electric field strength of the laser field, with $\epsilon_0$ and $c$ the permittivity of free space, and the speed of light in vacuum, respectively.

The dynamic polarizability of an individual $|nJM_J\rangle$ state can be calculated by summing the off-resonant single-photon dipole-allowed couplings induced by the laser field, of this state to all other non-degenerate bound and continuum states. For a state $|nJM_J\rangle$ lying at an  energy $E_{|nJM_J\rangle}$~\cite{Deissler08a}, 
\begin{eqnarray}
\alpha_{|nJM_J\rangle}(\omega_{\text{laser}}) &=& \frac{2}{\hbar}\sum_{|n'J'M'_J\rangle\neq|nJM_J\rangle}\Bigg[\frac{\Delta\omega}{\Delta\omega^{\,2}-\omega_{\text{laser}}^2}\nonumber\\
& & \hspace*{0.8cm}\times\, |\left\langle n'J'M'_J|\hat{\mu}|nJM_J\right\rangle|^2\Bigg],
\label{eq:pol}
\end{eqnarray}
where $\hbar=h/(2\pi)$ ($h$ is Planck's constant), $\Delta\omega=(E_{|n'J'M'_J\rangle}-E_{|nJM_J\rangle})/\hbar$, and $\hat{\mu}$ is the electric dipole operator.

In electronic states with low values of $n$, such as the initial $2\,^3$S$_1$ state in the experiments reported here, the electron is localized close to the positively charged ion core and high intensity far-off-resonant laser radiation in the visible region of the electromagnetic spectrum can readily polarize the atom, giving rise to energy shifts that are proportional to the laser intensity, $I_{\text{laser}}$. Strong ac Stark shifts observed for states with low values of $n$ result from a combination of the low density of electronic states in the surrounding energetic regions, and the general propensity for bound and continuum states to lie toward higher energies.

In Rydberg states with high values of $n$, the density of states in the surrounding energetic regions are high, and far-off-resonant visible laser radiation couples states lying higher in energy, and lower in energy, with similar coupling strengths. As a result the ac Stark energy shifts of high Rydberg states in strong high frequency (visible) laser fields are weak and, as demonstrated in the data presented here, can, to a reasonable approximation, be neglected when performing experiments with pulsed laser radiation with bandwidths exceeding 1~GHz.

The determination of the ac Stark shift of a state $|nJM_J\rangle$ using Equation~(\ref{eq:pol}) requires the calculation of the dipole matrix elements $\left\langle n'J'M'_J|\hat{\mu}|nJM_J\right\rangle$. These matrix elements can be calculated using the Wigner-Eckart theorem to  express them in terms of a reduced radial matrix element $\langle n'L' ||e\hat{r}||nL \rangle$ such that~\cite{zarebook}
\begin{widetext}
\begin{eqnarray}
\left\langle n'J'M'_J|\hat{\mu}|nJM_J\right\rangle &=& (-1)^{J+J'-M'_J+L'+S+1}\sqrt{(2J'+1)(2J+1)}\nonumber\\
& &\hspace*{1.5cm}\left( \begin{array}{ccc} J' & 1 & J \\ -M'_J & \Delta M_J & M_J\end{array} \right)\left\{ \begin{array}{ccc} L' & J' & S \\ J & L & 1\end{array} \right\}\langle n'L' ||e\hat{r}||nL \rangle,
\label{eq:dipmat}
\end{eqnarray}
\end{widetext}
where the curved and curly brackets represent the Wigner 3J and 6J symbols, respectively. In the particular case of the triplet states of the He atom of relevance here, the single electron in the 1s orbital has zero orbital angular momentum. Since this orbital lies 19.8~eV below the next nearest excited level to which it can be coupled by the laser field, in our treatment of the ac Stark shift of the 2\,$^3$S$_1$ state we consider only the single optically active electron with orbital angular momentum $\ell$, which is initially in the 2s orbital. Therefore $L\equiv\ell$ and the reduced radial matrix element $\langle n'L' ||e\hat{r}||nL \rangle\equiv\langle n'\ell' ||e\hat{r}||n\ell \rangle$.

From the expression in Equation~\ref{eq:dipmat} and using the Numerov method~\cite{zimmerman79a} to calculate the reduced radial matrix elements, together with previously published values for the $n$-dependent quantum defects, $\delta_{n\ell}$, of the triplet states of He ($\delta_{30\mathrm{s}}=0.2967$, $\delta_{30\mathrm{p}}=0.0683$, $\delta_{30\mathrm{d}}=0.0029$)~\cite{martin87a}, the polarizability of the 2\,$^3$S$_1$ state, in an electromagnetic field with a frequency $\omega_{\text{laser}}/(2\pi) = 5.738\times10^{14}$~Hz ($\equiv 522.48$~nm) -- half of the transition frequency associated with the energy difference between the 2s state and the 25d state -- was determined. In doing this the off-resonant couplings to all $n$p states with values of $n$ in the range from 2 to 20 were included. For this range of states, the calculated polarizability of $\alpha_{2\,^3\mathrm{S}_1}(\omega_{\mathrm{laser}})=-3.15\times10^{-39}$~Cm$^2$V$^{-1}$ (equivalent to a frequency shift of $-8.95~$GHz in a laser field of intensity $10^{13}$~Wm$^{-2}$) converged to better than one part per thousand.

Using this value for the dynamic polarizability of the 2\,$^3$S$_1$ state, the spectral lineshapes associated with the two-photon transitions studied experimentally were calculated. This was done by determining the ac Stark shift of the 2\,$^3$S$_1$ state of a spatially distributed ensemble of atoms interacting with the laser beam. The spatial dimensions of the three-dimensional ensemble matched the known shape of the supersonic beam (uniform distributions in $y$- and $z$-dimensions with width $400~\mu$m, much greater than the size of the laser beam in these dimensions, and a Gaussian distribution in the $x$-dimension with width $\sigma_x=1.5~$mm). The measured laser beam profile and convergence at the intersection region between it and the atomic beam (a Gaussian transverse laser profile with $\sigma_y\simeq\sigma_z\simeq70\mu$m and gradient $\text{d}\sigma_{y,z}/\text{d}x=-11.5 \mu$m(mm)$^{-1}$) were then used to generate a three-dimensional intensity distribution for each laser pulse energy, from which the ac Stark shifts could be calculated. The resulting ac Stark shifted transition frequency of each atom in the ensemble was then convolved with the measured bandwidth of the laser ($\sigma_L=\sqrt{2}~5.0~$GHz at the two-photon level), and the contributions from all atoms were then summed to determine the overall lineshape. Further details of the numerical procedure used to calculate the spectral lineshapes observed experimentally are provided in the Appendix.

\section{\label{results}Results}
\subsection{Photoexcitation to high Rydberg states}
\begin{figure*}[!]
\centerline{\includegraphics[width=1 \linewidth]{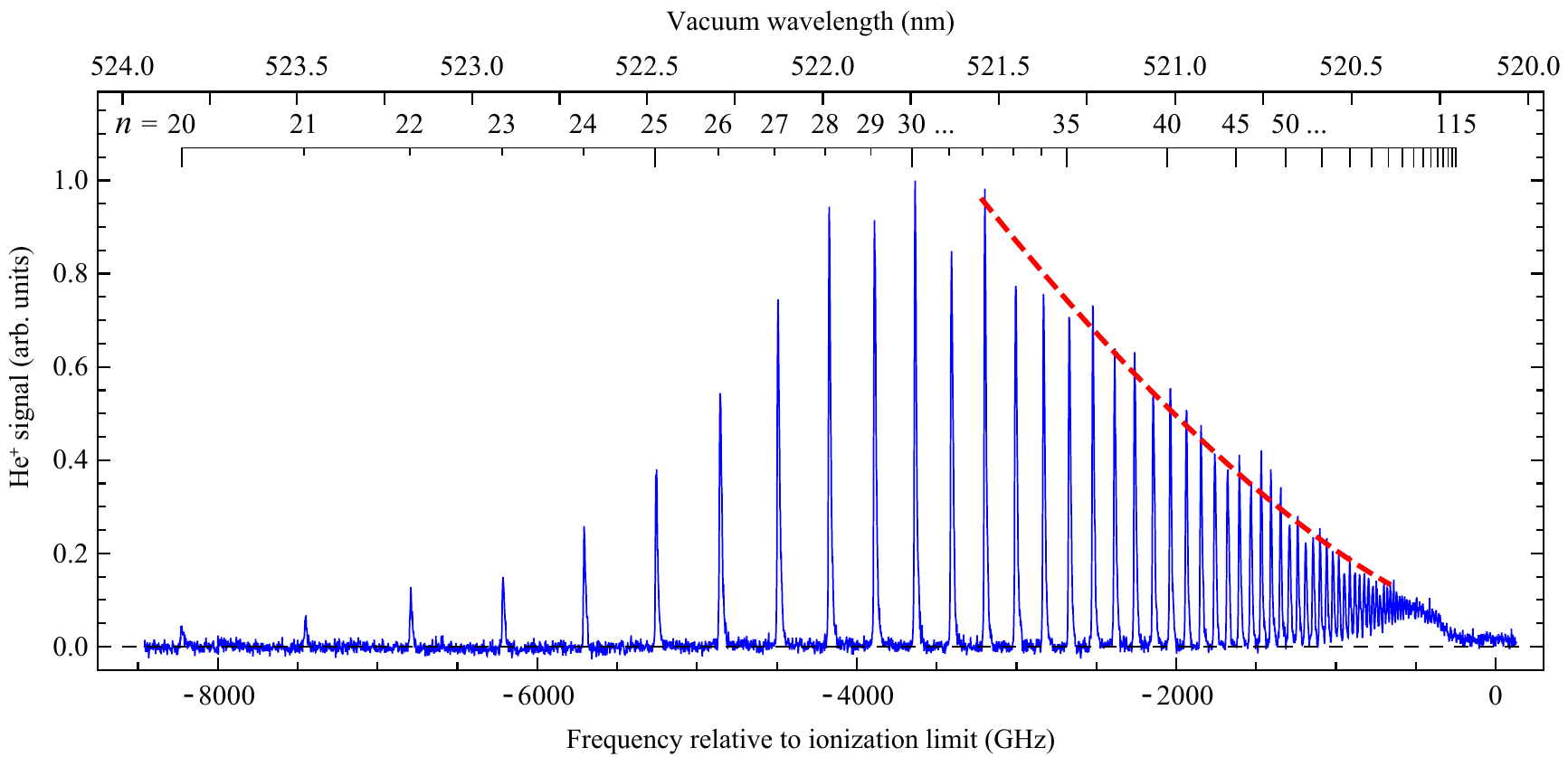}}
\caption{Spectrum of triplet $n$d Rydberg states of He with $|m_{\ell}|=2$ recorded following non-resonance-enhanced single-color two-photon excitation. The calculated line positions are indicated by the vertical bars at the top of the figure. The red dashed line represents a function the amplitude of which depends on $n^{-3}$ that has been fitted to the amplitudes of the spectral features with $32\leq n \leq 70$. The offset of the zero intensity level close to the ionization limit is a result of electrical fluctuations during the measurement. Careful measurements made with and without the laser beam present confirm that there is no $\text{He}^+$ signal recorded at the ionization limit.}
\label{fig:longscan}
\end{figure*}

\begin{figure}[!]
\centerline{\includegraphics[width=1.0 \linewidth]{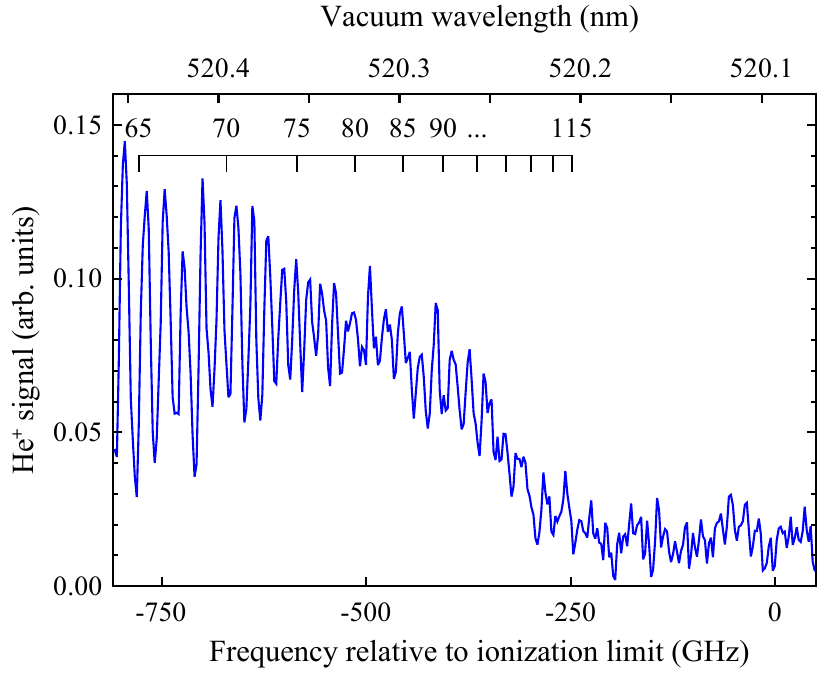}}
\caption{(Color online) The single-color two-photon $|m_{\ell}|=2$ spectrum encompassing transitions to states from $n=65$ up to the ionization limit. The spectral features are asymmetric, broadened to higher frequency by the ac Stark shift of the 2\,$^3{\mathrm S}_1$ state (see Sec. \ref{ACStarkExp}). The decrease in signal strength above $n=85$ is a result of electric field ionization in the fringe fields of the MCP between the photoexcitation region and detection region.}
\label{fig:specZoom}
\end{figure}
A single-color two-photon spectrum of $n$d Rydberg states of helium with $|m_{\ell}|=2$ is presented in Fig.~\ref{fig:longscan}. This spectrum was recorded by scanning the wavelength of the pulsed dye laser from 520~nm to 524~nm. This range covers two-photon transitions to Rydberg states ranging from $n=20$ up to the ionization limit. The highest Rydberg states observed are those with $n\simeq 100$, while states up to  approximately $n=75$ are clearly resolved (see Fig.~\ref{fig:specZoom}). In Figure \ref{fig:longscan} the calculated positions of the Rydberg states are indicated by the vertical bars above the spectrum and are in very good agreement with the positions of the experimentally recorded peaks. Careful inspection of Fig.~\ref{fig:longscan} reveals that each spectral feature is asymmetric, broadened to the high frequency side. This broadening is a  result of the ac Stark shift of the 2\,$^3{\mathrm S}_1$ state (see Section~\ref{acStarkSec}), and is discussed in more detail in Section~\ref{ACStarkExp}.

The overall shape of the spectrum presented in Fig.~\ref{fig:longscan} results from several competing effects. The reduction in the intensity of the spectral features from $n=30$ up to $n=85$ scales with $n^{-3}$ (see dashed curve), and is a result of the reduction in the transition dipole moments from the virtual intermediate state at the single photon level with increasing $n$. The cut-off in signal above $n=110$, below the ionization limit, is a result of electric-field ionization of the excited atoms in the fringe fields of the MCP detector before reaching the position in the detection region where they can be efficiently collected to contribute to the measured signal. The gradual increase in the intensity of the spectral features for values of $n$ from 20 to 30 arises from a combination of effects of increasing radiative lifetimes, transitions driven by blackbody radiation and the cut-off of electric field ionization  at low $n$ in the detection region of the apparatus. In the absence of blackbody radiation, the $\approx800~$Vcm$^{-1}$ pulsed electric field used for Rydberg atom ionization/detection is not strong enough to ionize atoms in states below $n=(F_0/9F)^{1/4}=29$, where $F_0=2R_{\text{He}}h c/e a_0=5.14\times10^9~$Vcm$^{-1}$, with $R_{\text{He}}$ the Rydberg constant for He. However, in the 60~$\mu$s flight-time of the atoms from the position of photoexcitation to the detection region, transitions from states with values of $n$ below 29 to those above permit detection of these lower $n$ states~\cite{farley81a}.

An expanded view of the portion of the two-photon spectrum above $n=65$ is displayed in Fig.~\ref{fig:specZoom}. From this data individual features up to approximately $n=75$ can be clearly identified. States with $n>75$ are not well resolved for two reasons. Firstly, at high $n$ the states are very closely spaced in energy, and for $n>94$ the separation of adjacent states becomes smaller than the experimental linewidth. This width,  $\simeq 20~$GHz (FHWM) as a result of ac Stark shifts (see Sec.~\ref{ACStarkExp}), is comparable to the level spacing for $n\simeq 74$. Secondly, these high-$n$ states possess very large electric dipole moments and weak stray electric fields in the excitation region will broaden the corresponding spectral features. For example, a field of only 0.1~Vcm$^{-1}$ will cause the $n=76$ state to broaden to $\sim2.5$~GHz. Finally, for states with $n\geq85$ the measured signal begins to decrease, reaching zero intensity at $n \simeq 110$. This cut-off is a result of electric field ionization in weak fringe electric fields of the MCP, between the excitation and detection regions. A stray electric field of $\sim10~$Vcm$^{-1}$ is sufficient to begin to ionize states with $n\approx85$ and strong enough to fully ionize states with $n>110$, where $F_{\text{ion}}=2F_0/(9 n^4)~$. Since the range of values of $n$ corresponds well to the extent of the cut-off region in Fig.~\ref{fig:specZoom}, we conclude that between the photoionization region and detection region atoms encounter fringe electric fields from the MCP of $\sim10~$Vcm$^{-1}$, precluding detection of states above $n\simeq 115$. This drop in signal at high $n$ also indicates that in the apparatus used here, $\text{He}^+$ ions formed by multiphoton ionization in the excitation region, or blackbody photoionization are not detected. This confirms that the recorded spectral features arise from atoms in long-lived Rydberg states and not from free ions.

\subsection{Effects of laser polarization}
\begin{figure}[t]
\centerline{\includegraphics[width=1 \linewidth]{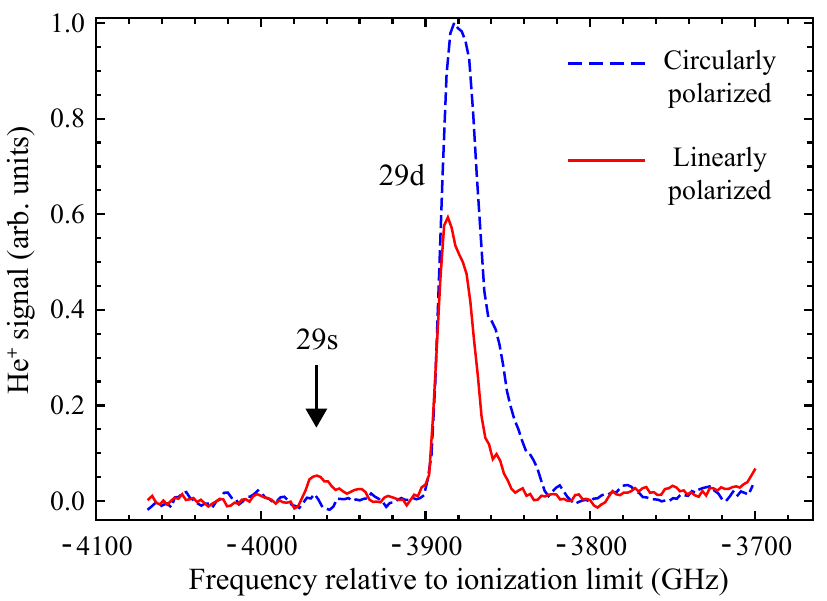}}
\caption{(Color online) The spectral region in the vicinity of the transition to the  29d state measured with circularly (blue, dashed curve) and linearly (red, continuous curve) polarized radiation. With circular polarized radiation only the transition to the 29d state is observed while with linear polarized radiation the transition to the 29s state is also seen.}
\label{fig:spscans}
\end{figure}
The spectrum in Fig.~\ref{fig:longscan} contains transitions to $n$d Rydberg states with $|m_{\ell}|=2$ for atoms excited using circularly polarized laser radiation. From the selection rules for electric dipole transitions, excitation with linearly polarized light leads to transitions for which $\Delta m_{\ell}=\pm1$. From the virtual intermediate state with $m_{\ell}=\pm1$ the second photon can therefore drive transitions either to Rydberg states with $m_{\ell}=0$, or with $|m_{\ell}|=2$. The latter transitions will populate $n$d-states with $|m_{\ell}|=2$, whereas the former will populate $n$s- and $n$d-states with $m_{\ell}=0$ [Fig.~\ref{fig:expSetup}(b)]. To confirm the $|m_{\ell}|$-selective character of the excited states, spectra are presented in Fig.~\ref{fig:spscans}, that were recorded close to $n=29$ using linearly and circularly polarized laser radiation. In the case of circularly polarized excitation, only the transition to the 29d state is observed, while in the case of linearly polarized excitation transitions to the short-lived 29s state, and the 29d state can be seen.

The transition to the 29s state occurs at a frequency $\sim80~$GHz below the transition to the 29d state because of the larger value of the 29s quantum defect ($\delta_{29s}=0.2967$, $\delta_{29d}=0.0029$~\cite{martin87a}). The reduced intensity of the transition to the 29d state in the case of excitation with linear polarized radiation is a result of the smaller angular integral associated with the transition moments to excited $n$d-states with $m_{\ell}=0$ than to $n$d-states with $|m_{\ell}|=2$. In the case here, the spectral intensity associated with the transition to the 29d states with $|m_{\ell}|=0$ and 2 deiven using linearly polarized radiation is expected to be 0.61 times that associated with the excitation of the 29d state with $|m_{\ell}|=2$ using circularly polarized radiation of the same intensity. As can be seen from the data in Fig.~\ref{fig:spscans}, this intensity ratio is consistent with that observed experimentally. The signal associated with the transition to the 29s state is weaker because of a smaller transition dipole moment than that to the 29d state and because of its shorter fluorescence lifetime. This reduced lifetime makes it difficult to compare the observed relative intensity of this transition to that expected theoretically.

\subsection{ac Stark shift}\label{ACStarkExp}
As discussed in Sec \ref{acStarkSec} the intense laser radiation employed to drive the non-resonance-enhanced single-color two-photon transitions presented here causes the $2\,^3\mathrm{S}_1$ state to shift to lower energy with increasing intensity, this shifts the 2s-$n$d transitions to higher frequency. The spatial spread of the atoms in the supersonic beam,which is much larger than the laser beam waist, leads to an asymmetric broadening of the observed transitions, since the atom cloud covers a volume over which there is a considerable gradient in the laser intensity. Atoms that are in the center of the laser beam, where the intensity is greatest,  experience a large ac Stark shift, while those at the edge of the beam experience only a small shift. Since the atoms that are subjected to the greatest intensities are also the most likely to be excited to a Rydberg state, this would indicate that for a collimated laser beam the spectral features would be shifted towards the frequency associated with the greatest ac Stark shift. However, the recorded spectra show features with maxima close to the unperturbed line-position, with a long tail towards higher frequency. This line shape is a result of the intensity gradient of the converging laser beam, which reaches a focus beyond the atomic beam axis. Atoms that interact with the laser beam at a position close to the focus can experience strong ac Stark shifts, while atoms located further from the focus experience weaker shifts. The ensemble of atoms located further from the focus experiences a smaller range of ac Stark shifts, causing the overall signal from these atoms to `bunch up' close to the unperturbed transition frequency, whereas closer to the focus atoms experience a wider range of intensities, causing the signal from these atoms to be spread out in frequency.
\begin{figure}[t]
\centerline{\includegraphics[width=1 \linewidth]{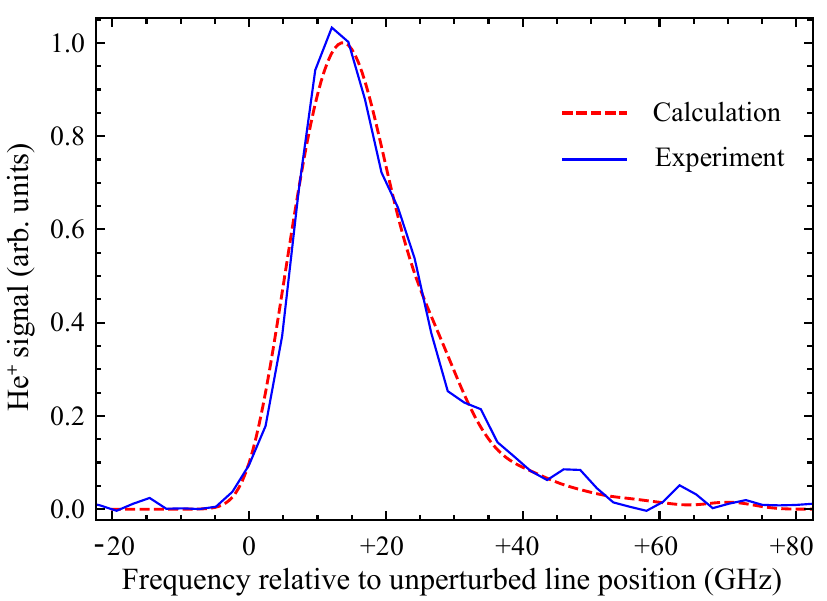}}
\caption{(Color online) Experimentally measured spectrum of the transition to the 26d state, recorded with a laser energy of 9.5~mJ per pulse focussed to a beam width of $\sigma_x\simeq\sigma_y\simeq70~\mu$m resulting in a spatially averaged peak laser intensity of $I_{\mathrm{laser}}=2.3\times10^{13}~$Wm$^{-2}$ (blue, continuous curve), and the calculated line shape (red, dashed curve)}
\label{fig:acStarkN26}
\end{figure}

The cumulative effect of the three-dimensional ensemble of atoms interacting with the converging laser beam is a spectral line with a maximum close to the unperturbed transition frequency, and a long tail towards higher frequencies. An example of this is shown in Fig.~\ref{fig:acStarkN26} where the spectral feature corresponding to the transition to the $26$d state is shown. In this figure the frequency is displayed relative to the unperturbed $26$d transition frequency. The transition was measured with a spatially averaged peak laser intensity on the atomic beam axis of $2.3\times10^{13}~$Wm$^{-2}$, corresponding to a maximal ac Stark shift of $21~$GHz. The data show a strong peak located $\sim12~$GHz above the unperturbed frequency, with a FWHM of 19~GHz and a long tail extending to $\sim60~$GHz. This long tail results from the contribution of atoms passing through the beam closer to the focus, where the local peak intensity is $7.3\times10^{13}~$Wm$^{-2}$, and the maximum ac Stark shift is 65~GHz. Also shown is the result of a numerical calculation that accounts for the laser beam profile and the shape of the atomic beam, and is in excellent agreement with the experimental data.

\begin{figure}[t]
\centerline{\includegraphics[width=1 \linewidth]{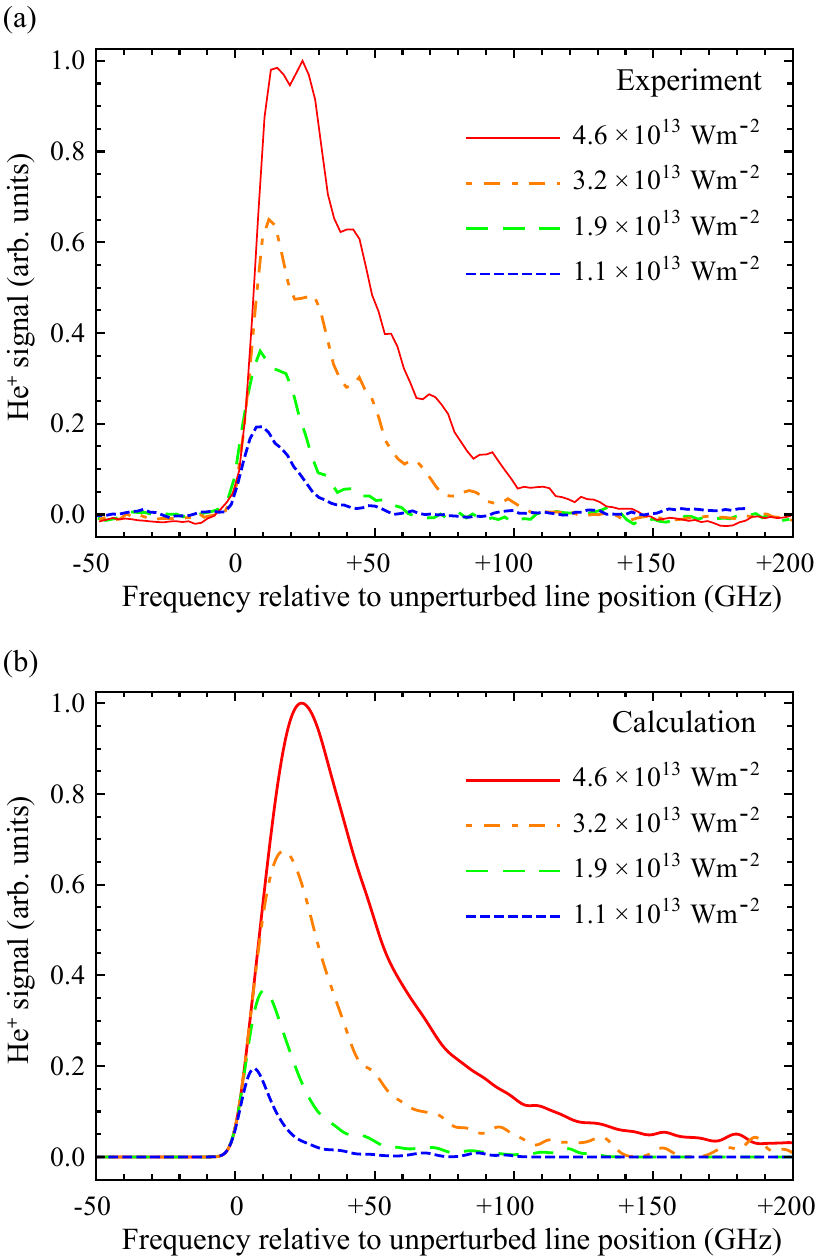}}
\caption{(Color online) (a) Measured and (b) calculated spectra of the transition to the 29d state for peak laser intensity $I_{\text{laser}}=4.6\times10^{13}~$Wm$^{-2}$ (red, continuous curve), $3.2\times10^{13}~$Wm$^{-2}$ (orange, dashed-dotted curve), $1.9\times10^{13}~$Wm$^{-2}$ (green, long-dashed curve), and $1.1\times10^{13}~$Wm$^{-2}$ (blue, short-dashed curve).}
\label{fig:sigVSenergy}
\end{figure}

Figure \ref{fig:sigVSenergy}(a) shows spectra close to the transition to the 29d state recorded with a range of spatially averaged peak laser intensities. With a low intensity  on the atomic beam axis of $I_{\text{laser}}=1.1\times10^{13}~$Wm$^{-2}$ the spectral profile is slightly asymmetric, and is centered close to the unperturbed transition frequency. As the intensity is increased the profile becomes gradually more asymmetric, with the tail shifting out to higher frequencies. The maximum also moves gradually to higher frequencies, but remains close to the unperturbed transition frequency. Figure \ref{fig:sigVSenergy}(b) shows calculated line profiles which are in very good agreement with the experimental data, confirming the accuracy of the numerical procedure used to model the excitation process. In the spectra recorded at high laser intensities, the fluctuations in the He${^+}$ ion signal exceeds the shot-to-shot fluctuations in the experiment. The observed fluctuations (most clearly displayed in Fig.~\ref{fig:sigVSenergy}(a) in the data recorded with $I_{\text{laser}}=4.6\times 10^{13}~$Wm$^{-2}$) appear to result from interference between different pathways in the excitation process, and will be investigated in more detail in the future.

\begin{figure}[t]
\centerline{\includegraphics[width=1 \linewidth]{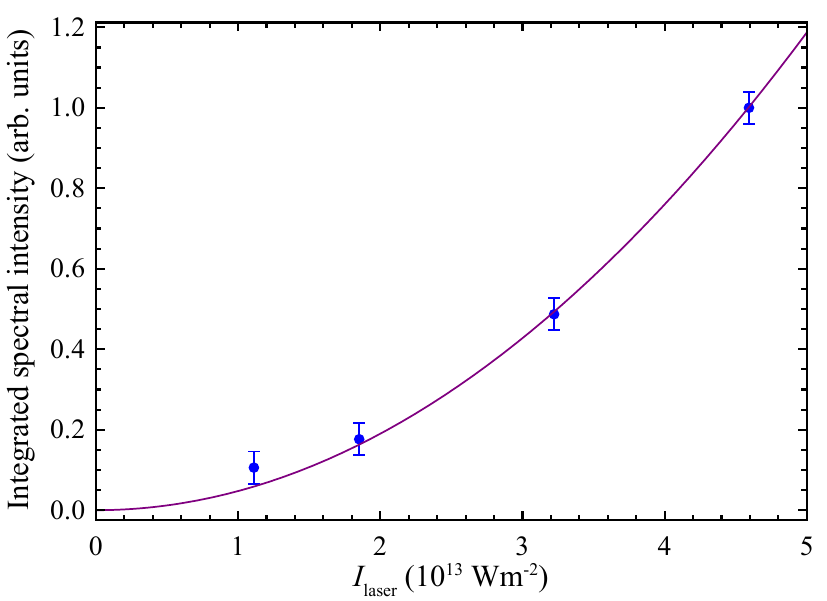}}
\caption{(Color online) Integrated spectral signal for the measured transitions to the 29d state shown in Fig.~\ref{fig:sigVSenergy} as a function of the peak laser intensity along the atomic beam axis, $I_{\mathrm{laser}}$. Also shown is a quadratic function that has been fit to the experimental data (purple continuous curve).}
\label{fig:intVSenergy}
\end{figure}

The spectral intensity of the non-resonance-enhanced two-photon transitions studied here are expected to be proportional to $I_{\mathrm{laser}}^2$ because of the intensity-dependence of the two-photon process. However, the features in the recorded spectra are broadened by the intensity-dependent ac Stark shift of the $2\,^3\mathrm{S}_1$ state. Therefore the integral of the spectral intensity is the quantity dependent upon $I_{\mathrm{laser}}^2$, and not simply the spectral amplitude. This can be seen in Fig.~\ref{fig:intVSenergy} where the integral of the spectral intensity over the frequency range encompassed in Fig.~\ref{fig:sigVSenergy} is displayed as a function of the spatially averaged peak laser intensity on the atomic beam axis, together with a quadratic function which confirms the $I_{\text{laser}}^2$  dependence of the integrated signal.

\begin{figure}[t]
\centerline{\includegraphics[width=1 \linewidth]{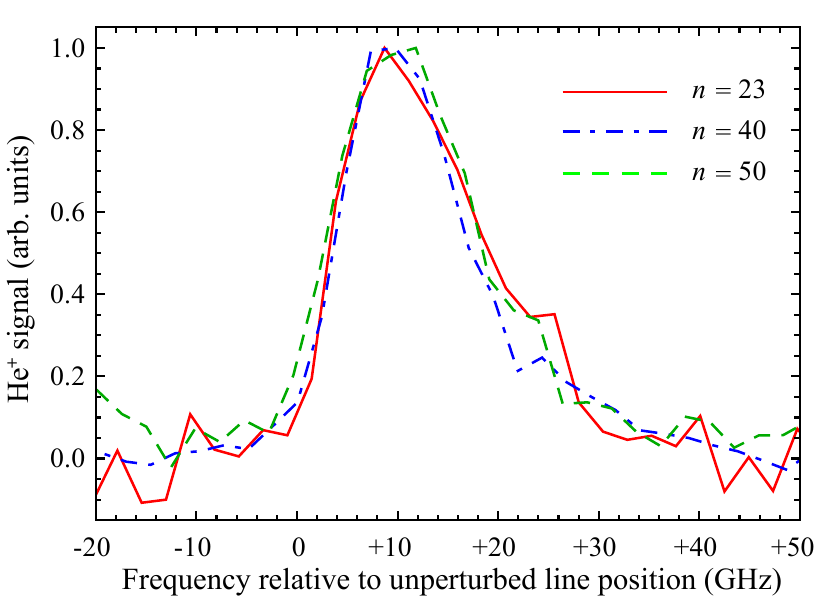}}
\caption{(Color online) Measured spectral features associated with the $n=23$ (red, continuous curve), 40 (blue, dashed-dotted curve) and 50 (green, dashed curve) Rydberg states, scaled to equal amplitude for ease of comparison. The spectral features were recorded with the same laser intensity, and clearly demonstrate the $n$-independence of the spectral line profile.}
\label{fig:peakWidthsN}
\end{figure}

Because the ac Stark shift is a feature of the interaction of the 2\,$^3$S$_1$ state with the laser field, and does not affect the Rydberg states, similar asymmetric lineshapes are seen for all recorded two-photon transitions. This is also the reason why all the spectral features in Fig.~\ref{fig:longscan} are shifted slightly to higher frequency from the calculated line positions. The peak widths are derived from the overlap of the atoms with the laser beam (as described in the Appendix), and are thus independent of the principal quantum number of the Rydberg states. This is illustrated in Fig.~\ref{fig:peakWidthsN} which shows the $n = 23$, 40 and 50 spectral features (scaled to equal amplitude for ease of comparison) overlaid, clearly demonstrating the $n$-independence of the line shape.

\subsection{Electric field}
The application of a homogeneous dc electric field in the excitation region mixes Rydberg states with the same value of $m_{\ell}$ but with values of $l$ differing by $\pm1$.
\begin{figure*}[t]
\centerline{\includegraphics[width=.6 \linewidth]{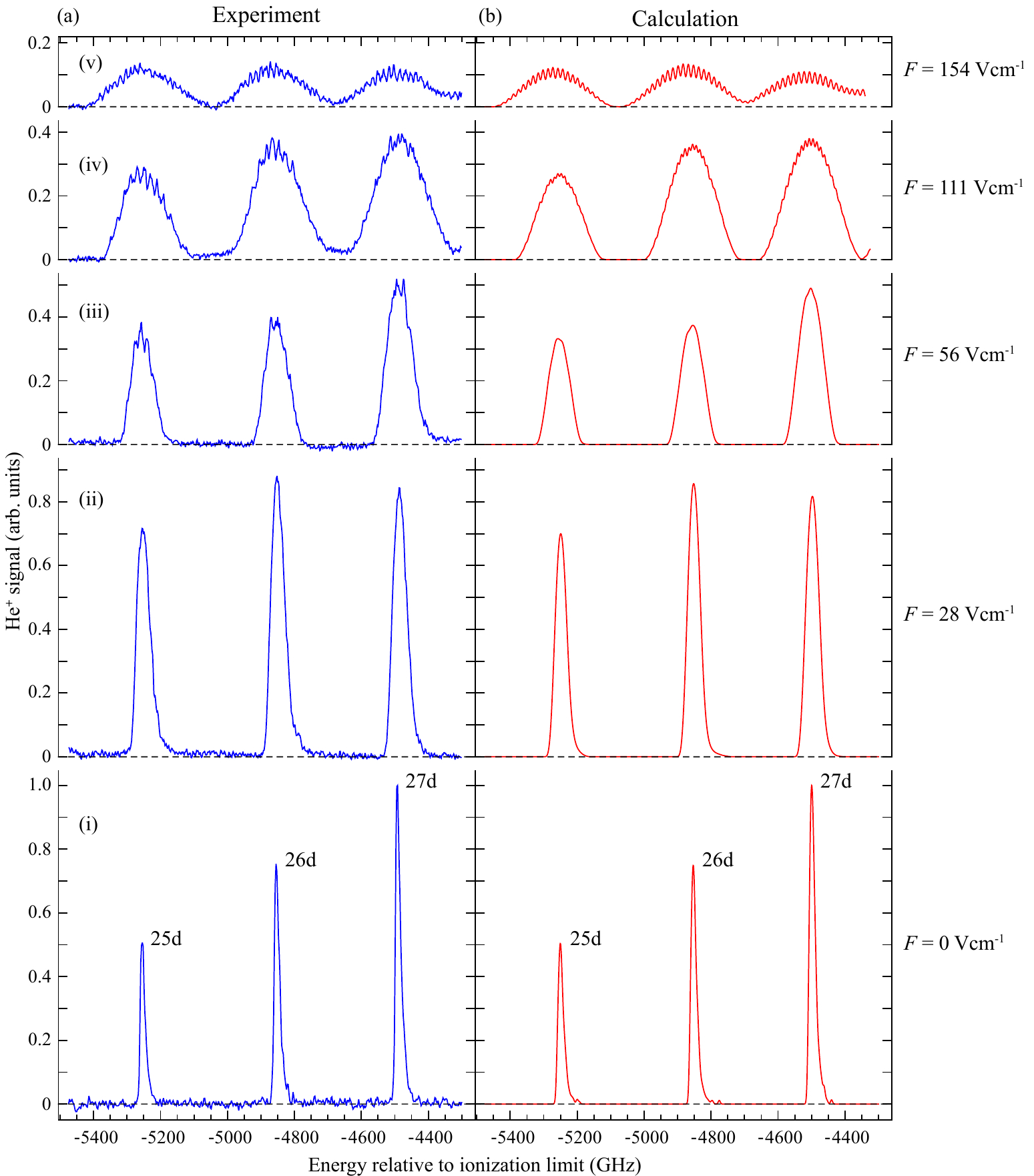}}
\caption{(a) Spectra of $|m_{\ell}|=2$ Rydberg states of He with $n=25$--27, recorded with a spatially averaged peak laser intensity of $I_{\text{laser}}=2.3\times10^{13}$Wm$^{-2}$ in the presence of dc electric fields of (i) 0~Vcm$^{-1}$, (ii) 28~Vcm$^{-1}$, (iii) 56~Vcm$^{-1}$, (iv) 111~Vcm$^{-1}$ and (v) 154~Vcm$^{-1}$.(b) Calculated spectra for the same conditions as those of the experimental data in (a).}
\label{fig:StarkPlotN}
\end{figure*}
As the applied electric field is increased the resulting Stark states in a given $n$-manifold separate in energy. Figure \ref{fig:StarkPlotN}(a) shows measured spectra of the transitions to the $n=25$, 26 and 27 states in a range of dc electric fields applied in the excitation region. In zero field the lines are asymmetric, with widths of $\sim20$~GHz (FWHM) resulting from the ac Stark shift (Sec. \ref{ACStarkExp}). As the field is increased to 28~Vcm$^{-1}$ the lines start to broaden symmetrically as the Stark states begin to separate in energy. Increasing the field further to 56~Vcm$^{-1}$ the energy splittings of the Stark states further increase, with the features becoming symmetric as the dc Stark shifts in the Rydberg states dominate the ac Stark shifts of the 2\,$^3$S$_1$ state. With this applied field strength the Stark states begin to be resolved.

\begin{figure}[t]
\centerline{\includegraphics[width=1 \linewidth]{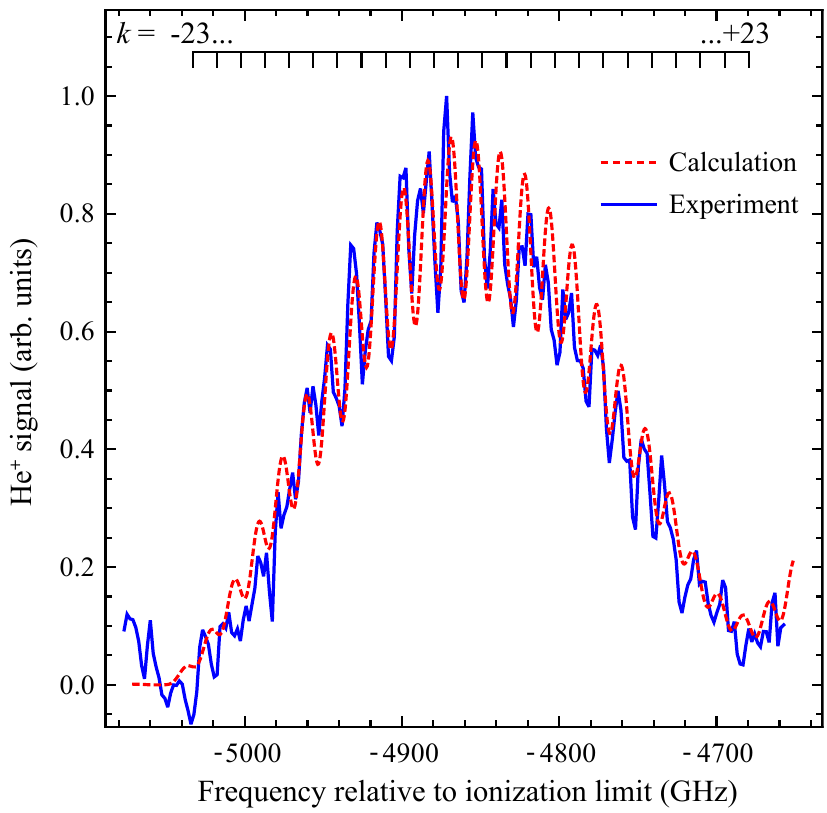}}
\caption{(Color online) Spectrum of the $n=26$ Stark manifold of He in a dc electric field of 154~Vcm$^{-1}$ (blue continuous curve), shown with the calculated spectrum (red, dashed curve). Each of the 24 Stark states are labelled with the index $k$ which ranges from $-(n-|m_{\ell}|-1)$ to $+(n-|m_{\ell}|-1)$ in steps of 2. The lowest-energy Stark states of the $n=27$ manifold are visible toward the high-frequency end of the spectrum.}
\label{fig:StarkPlotZoom}
\end{figure}

As the electric field is increased further the dc Stark shifts continue to increase and the spectral features become more symmetric. In a field of 111~Vcm$^{-1}$, the Stark state resolution improves and the structure is clearer. With the maximum applied field of 154~Vcm$^{-1}$ the Stark states are clearly separated with each individual state identifiable. Note that in this field the $n=26$ and $n=27$ Stark manifolds are sufficiently broad that they overlap. This field is greater than the Inglis-Teller field of $\sim144~$Vcm$^{-1}$ for $n=26$~\cite{gallagher94a}. Fig.~\ref{fig:StarkPlotN}(b) shows calculations corresponding to the experimental data in Fig.~\ref{fig:StarkPlotN}(a). These spectra were calculated by convolving the calculated line profile arising from the ac Stark shift of the $2\,^3\mathrm{S}_1$ state with the spectral intensities of the transitions to each individual Stark state determined by diagonalization of the Hamiltonian matrix~\cite{zimmerman79a}. Fig.~\ref{fig:StarkPlotZoom} shows the spectrum at $n=26$ in a field of 154~Vcm$^{-1}$ in more detail. In this spectrum the 24 Stark states can be clearly identified. Also shown is the calculated spectrum which agrees well with the experimental data. The outer Stark states are not completely resolved because they are more susceptible to electric field inhomogeneities in the excitation region. Since the transition strengths to these states is weakest their spectral appearance is also more susceptible to shot-to-shot laser intensity variations, which slightly wash out their structure.

These measurements show that by applying a dc electric field in the excitation region it is possible to split the Stark states sufficiently to allow selective excitation using this non-resonance-enhanced two-photon scheme. The limiting factor in the resolution is the ac Stark shift of the initial state.

\section{Application to Doppler-free photoexcitation of Rydberg states of positronium}

In addition to being of importance for precision spectroscopic studies and the preparation of long-lived high-angular-momentum Rydberg states, non-resonance enhanced Doppler-free two-photon excitation schemes are expected to be particularly well suited to the photoexcitation of Rydberg states of positronium (Ps) with high spectral resolution.

Positronium atoms are most efficiently created by implanting positrons into solid state materials. Various mechanisms for Ps formation exist, which are either thermal in nature~\cite{Mills79}, or rely on other energy transfer processes~\cite{cassidy2010}, and typically produce atoms with velocities on the order of $10^5$~ms$^{-1}$. This leads to very large Doppler broadenings of single photon transitions (e.g., increasing the 50~MHz natural width of the 1s--2p transition to $\sim 500~$GHz~\cite{cassidy12a}). Such large Doppler widths preclude the selective excitation of individual Ps Rydberg-Stark states using previously demonstrated two-color two-photon excitation schemes~\cite{Jones14}. However, in order to overcome some of the challenges associated with performing gravitational~\cite{mills02a} or spectroscopic~\cite{Fee93} measurements on Ps it is desirable to photoexcite them to well-defined Rydberg-Stark states with long radiative lifetimes, and large electric dipole moments, so that they can be guided, decelerated and manipulated using inhomogeneous electric fields~\cite{vliegen04a,vliegen2006,hogan08a,hogan11a}. By implementing two-photon excitation schemes of the kind demonstrated here, but using two counter-propagating laser beams in a Doppler-free configuration~\cite{Haensch75a}, a spectral resolution limited by the combination of the laser bandwidth and the ground state ac Stark shift will be achievable. This approach would provide sufficient resolution to selectively prepare individual Rydberg-Stark states of positronium (as demonstrated for He in Fig.~\ref{fig:StarkPlotZoom}).

In the experiments reported here, using circularly polarized laser radiation at a wavelength of $\sim520~$nm, with a spatially averaged peak intensity of $\sim5\times10^{13}~$Wm$^{-2}$, approximately 10\% (determined by comparison with the measured He$^+$ ion signal recorded by single photon excitation at saturation) of the metastable helium atoms that interact with the laser beam were excited from the 2\,$^3\text{S}_1$ state to $n$d Rydberg states, close to $n=25$. In these non-resonance-enhanced single-color two-photon excitation processes the transition rates depend on the sum of the off-resonant (electric dipole allowed) couplings at the single-photon level, between the initial state and all available non-degenerate bound and continuum intermediate states, and the further off-resonant couplings from these intermediate states to the excited Rydberg state~\cite{Haas2006}. By summing the contributions of all far off-resonant intermediate states up to $n=50$ we have determined the approximate transition rate from the 2\,$^3\text{S}_1$ state to the triplet 25d state in He to be $2\times10^{-5}~$Hz/(Wm$^{-2}$).

Using a Doppler-free single-color two-photon excitation scheme to drive transitions from the 1s ground state of ortho-Ps to $n$d Rydberg states will require circularly polarized laser radiation at $\sim 364~$nm. From the exact calculations for two-photon transitions in hydrogenic systems by Haas~et~al.~\cite{Haas2006} the corresponding transition rate for the 1s$\rightarrow$25d transition in Ps is also $\sim 2\times10^{-5}~$Hz/(Wm$^{-2}$). The similarity of this number to the corresponding rate for the two-photon transitions in He reported here indicates that similar excitation efficiencies can be expected in Ps. In addition, the frequency-dependent ground state polarizability of Ps, calculated using the methods outlined in Sec. \ref{acStarkSec} for laser radiation with a photon energy appropriate for driving the Doppler-free two-photon 1s$\rightarrow$25d transition is $+1.4\times10^{-39}~$Cm$^2$V$^{-1}$ (equivalent to a frequency shift of +4~GHz in a laser field of intensity $10^{13}~$Wm$^{-2}$), with a magnitude of approximately half that of the $2\,^3\mathrm{S}_1$ state in He. It is expected that this lower dipole polarizability and hence reduced ac Stark energy shift at a given laser intensity will also enhance the excitation efficiency in Ps.

The laser intensities used to drive single-color two-photon transitions in He were achieved by focussing the output of a pulsed dye laser to a width of $\sigma\sim70~\mu$m. While it is possible to employ greater laser powers using commercially-available dye laser systems, a practical upper limit to the dimension of the beam size required to obtain the appropriate intensity is expected to be $\sigma\sim200~\mu$m, corresponding to a FWHM of $\sim0.5~$mm. The production of a $\sim1~$mm diameter positron (and hence also Ps) pulse from an accumulator can be achieved with moderate magnetic fields of 500~G~\cite{Cassidy2006}. For experiments in which such fields are acceptable this means that approximately half of the available atoms could be addressed, and $\sim5\%$ excited. However, because Ps velocities are generally high, ionization of Rydberg states  by motionally induced electric fields can lead to significant losses~\cite{cassidy12a,hogan13a}. For magnetic fields of 50~G, such losses are largely mitigated, but then the positron beam size will be $\sim 3~$mm, and the overlap with a 0.5~mm FWHM laser beam would be about 15\%, leading to an overall excitation efficiency of 1.5\%. A typical Ps pulse in an apparatus operating at 1~Hz would contain $\sim10^6$ atoms, so that more than $10^4$ Rydberg atoms could be produced per second.

An alternative approach is to remoderate the positron beam~\cite{Mills1980}. In this arrangement the positrons would be extracted from the trap magnetic field completely, and focussed into a solid material (the remoderator). Non-conservative interactions in the remoderator make it possible to circumvent phase-space  limitations, and the re-emitted positron beam brightness may be considerably enhanced. For typical trap parameters this would result in a loss of $\sim 90\%$ of the particles, but with a spot size of less than $100~\mu$m. In this case the initial $10^6$ positrons would again result in $\sim10^4$ Rydberg atoms, but this time in zero magnetic field, and with a spatial spread smaller than the laser beam size, so that significantly smaller laser pulse energies ($\sim$1~mJ) would be required.

A further reduction in the required laser energy can be achieved by using a laser with a reduced bandwidth. Seeding a pulsed dye laser amplifier, with, for example a narrow-band continuous wave (cw) diode laser tuned to the desired fundamental wavelength, will result in a reduction in the laser bandwidth by a factor of $\sim20$. Using such a transform-limited pulse will reduce the energy required for a similar excitation efficiency by an order of magnitude. The use of transform-limited laser radiation and a brightness-enhanced re-moderated positron source could reduce the energy required to drive two-photon transitions in Ps to $\sim 100~\mu$J per pulse.

\section{\label{concl}Conclusions}
Non-resonance-enhanced single-color two-photon excitation to long-lived Rydberg states in He has been demonstrated by driving transitions from the metastable $2\,^3\text{S}_{1}$ state to Rydberg states in the range from $n=20$ up to $n\simeq 100$. The effect of the intensity and polarization of the laser radiation on the excitation efficiency and spectral profiles has been investigated in detail. The spectral features in the recorded spectra were observed to be broader than the bandwidth of the laser, and asymmetric with a long tail to high frequency. It has been shown by comparison of the experimental data with the results of numerical calculations that this is a result of the ac Stark shift of the $2\,^3\text{S}_{1}$ state in the laser field, which non-resonantly couples the $2\,^3\text{S}_{1}$ state to other bound states. The polarizability of the $2\,^3\text{S}_{1}$ state was calculated by summing all off-resonant couplings to $n$p states in the range from $n=2$ up to 20. With detailed measurements of the laser beam profile this allowed the calculation of the spectral lineshapes, which were found to be in excellent agreement with those recorded experimentally.

Photoexcitation in dc electric fields was also performed, and with a field strength of 154~Vcm$^{-1}$ individual Rydberg-Stark states were clearly identified. This selective population of individual Rydberg-Stark states in a non-resonance-enhanced multi-photon photoexcitation scheme as demonstrated is important for the preparation of atoms and molecules in long-lived high-angular-momentum Rydberg states for deceleration and trapping experiments using inhomogeneous electric fields.

In particular, the application of similar schemes to Ps has been discussed. Based on the measured excitation rates in He under experimental conditions similar to those expected in experiments with Ps, we conclude that it is feasible to excite ground state Ps atoms to Rydberg states with $n\simeq25$ in Doppler-free two-photon transitions with sufficient resolution to address individual Stark states. This will enable new experiments designed to control Rydberg Ps atoms with inhomogeneous electric fields via their large electric dipole moments.

\begin{acknowledgments}
This work has been supported financially by the Department of Physics and Astronomy, and the Faculty of Mathematical and Physical Sciences at University College London, and by the Engineering and Physical Sciences Research Council under grant number EP/K028774/1. We are grateful to the Engineering and Physical Sciences Research Council Laser Loan Pool (loan number 13250005) for the loan of the laser system used in this research.
\end{acknowledgments}

\bibliography{library}

%merlin.mbs apsrev4-1.bst 2010-07-25 4.21a (PWD, AO, DPC) hacked
%Control: key (0)
%Control: author (8) initials jnrlst
%Control: editor formatted (1) identically to author
%Control: production of article title (-1) disabled
%Control: page (0) single
%Control: year (1) truncated
%Control: production of eprint (0) enabled
\begin{thebibliography}{38}%
\makeatletter
\providecommand \@ifxundefined [1]{%
 \@ifx{#1\undefined}
}%
\providecommand \@ifnum [1]{%
 \ifnum #1\expandafter \@firstoftwo
 \else \expandafter \@secondoftwo
 \fi
}%
\providecommand \@ifx [1]{%
 \ifx #1\expandafter \@firstoftwo
 \else \expandafter \@secondoftwo
 \fi
}%
\providecommand \natexlab [1]{#1}%
\providecommand \enquote  [1]{``#1''}%
\providecommand \bibnamefont  [1]{#1}%
\providecommand \bibfnamefont [1]{#1}%
\providecommand \citenamefont [1]{#1}%
\providecommand \href@noop [0]{\@secondoftwo}%
\providecommand \href [0]{\begingroup \@sanitize@url \@href}%
\providecommand \@href[1]{\@@startlink{#1}\@@href}%
\providecommand \@@href[1]{\endgroup#1\@@endlink}%
\providecommand \@sanitize@url [0]{\catcode `\\12\catcode `\$12\catcode
  `\&12\catcode `\#12\catcode `\^12\catcode `\_12\catcode `\%12\relax}%
\providecommand \@@startlink[1]{}%
\providecommand \@@endlink[0]{}%
\providecommand \url  [0]{\begingroup\@sanitize@url \@url }%
\providecommand \@url [1]{\endgroup\@href {#1}{\urlprefix }}%
\providecommand \urlprefix  [0]{URL }%
\providecommand \Eprint [0]{\href }%
\providecommand \doibase [0]{http://dx.doi.org/}%
\providecommand \selectlanguage [0]{\@gobble}%
\providecommand \bibinfo  [0]{\@secondoftwo}%
\providecommand \bibfield  [0]{\@secondoftwo}%
\providecommand \translation [1]{[#1]}%
\providecommand \BibitemOpen [0]{}%
\providecommand \bibitemStop [0]{}%
\providecommand \bibitemNoStop [0]{.\EOS\space}%
\providecommand \EOS [0]{\spacefactor3000\relax}%
\providecommand \BibitemShut  [1]{\csname bibitem#1\endcsname}%
\let\auto@bib@innerbib\@empty
%</preamble>
\bibitem [{\citenamefont {Gallagher}(1994)}]{gallagher94a}%
  \BibitemOpen
  \bibfield  {author} {\bibinfo {author} {\bibfnamefont {T.~F.}\ \bibnamefont
  {Gallagher}},\ }\href@noop {} {\emph {\bibinfo {title} {Rydberg Atoms}}}\
  (\bibinfo  {publisher} {Cambridge University Press, Cambridge, U.K.},\
  \bibinfo {year} {1994})\BibitemShut {NoStop}%
\bibitem [{\citenamefont {Liu}\ \emph {et~al.}(2009)\citenamefont {Liu},
  \citenamefont {Salumbides}, \citenamefont {Hollenstein}, \citenamefont
  {Koelemeij}, \citenamefont {Eikema}, \citenamefont {Ubachs},\ and\
  \citenamefont {Merkt}}]{liu09a}%
  \BibitemOpen
  \bibfield  {author} {\bibinfo {author} {\bibfnamefont {J.}~\bibnamefont
  {Liu}}, \bibinfo {author} {\bibfnamefont {E.~J.}\ \bibnamefont {Salumbides}},
  \bibinfo {author} {\bibfnamefont {U.}~\bibnamefont {Hollenstein}}, \bibinfo
  {author} {\bibfnamefont {J.~C.~J.}\ \bibnamefont {Koelemeij}}, \bibinfo
  {author} {\bibfnamefont {K.~S.~E.}\ \bibnamefont {Eikema}}, \bibinfo {author}
  {\bibfnamefont {W.}~\bibnamefont {Ubachs}}, \ and\ \bibinfo {author}
  {\bibfnamefont {F.}~\bibnamefont {Merkt}},\ }\href {\doibase
  http://dx.doi.org/10.1063/1.3120443} {\bibfield  {journal} {\bibinfo
  {journal} {J. Chem. Phys.}\ }\textbf {\bibinfo {volume} {130}},\ \bibinfo
  {eid} {174306} (\bibinfo {year} {2009})}\BibitemShut {NoStop}%
\bibitem [{\citenamefont {Paul}\ \emph {et~al.}(2009)\citenamefont {Paul},
  \citenamefont {Liu},\ and\ \citenamefont {Merkt}}]{paul09a}%
  \BibitemOpen
  \bibfield  {author} {\bibinfo {author} {\bibfnamefont {{\relax Th}.~A.}\
  \bibnamefont {Paul}}, \bibinfo {author} {\bibfnamefont {J.}~\bibnamefont
  {Liu}}, \ and\ \bibinfo {author} {\bibfnamefont {F.}~\bibnamefont {Merkt}},\
  }\href {\doibase 10.1103/PhysRevA.79.022505} {\bibfield  {journal} {\bibinfo
  {journal} {Phys. Rev. A}\ }\textbf {\bibinfo {volume} {79}},\ \bibinfo
  {pages} {022505} (\bibinfo {year} {2009})}\BibitemShut {NoStop}%
\bibitem [{\citenamefont {Sa\ss{}mannshausen}\ \emph
  {et~al.}(2013)\citenamefont {Sa\ss{}mannshausen}, \citenamefont {Merkt},\
  and\ \citenamefont {Deiglmayr}}]{sassmannshausen13a}%
  \BibitemOpen
  \bibfield  {author} {\bibinfo {author} {\bibfnamefont {H.}~\bibnamefont
  {Sa\ss{}mannshausen}}, \bibinfo {author} {\bibfnamefont {F.}~\bibnamefont
  {Merkt}}, \ and\ \bibinfo {author} {\bibfnamefont {J.}~\bibnamefont
  {Deiglmayr}},\ }\href {\doibase 10.1103/PhysRevA.87.032519} {\bibfield
  {journal} {\bibinfo  {journal} {Phys. Rev. A}\ }\textbf {\bibinfo {volume}
  {87}},\ \bibinfo {pages} {032519} (\bibinfo {year} {2013})}\BibitemShut
  {NoStop}%
\bibitem [{\citenamefont {Gleyzes}\ \emph {et~al.}(2007)\citenamefont
  {Gleyzes}, \citenamefont {Kuhr}, \citenamefont {Guerlin}, \citenamefont
  {Bernu}, \citenamefont {Deleglise}, \citenamefont {Busk~Hoff}, \citenamefont
  {Brune}, \citenamefont {Raimond},\ and\ \citenamefont
  {Haroche}}]{gleyzes06a}%
  \BibitemOpen
  \bibfield  {author} {\bibinfo {author} {\bibfnamefont {S.}~\bibnamefont
  {Gleyzes}}, \bibinfo {author} {\bibfnamefont {S.}~\bibnamefont {Kuhr}},
  \bibinfo {author} {\bibfnamefont {C.}~\bibnamefont {Guerlin}}, \bibinfo
  {author} {\bibfnamefont {J.}~\bibnamefont {Bernu}}, \bibinfo {author}
  {\bibfnamefont {S.}~\bibnamefont {Deleglise}}, \bibinfo {author}
  {\bibfnamefont {U.}~\bibnamefont {Busk~Hoff}}, \bibinfo {author}
  {\bibfnamefont {M.}~\bibnamefont {Brune}}, \bibinfo {author} {\bibfnamefont
  {J.-M.}\ \bibnamefont {Raimond}}, \ and\ \bibinfo {author} {\bibfnamefont
  {S.}~\bibnamefont {Haroche}},\ }\href {\doibase 10.1038/nature05589}
  {\bibfield  {journal} {\bibinfo  {journal} {Nature}\ }\textbf {\bibinfo
  {volume} {446}},\ \bibinfo {pages} {297} (\bibinfo {year}
  {2007})}\BibitemShut {NoStop}%
\bibitem [{\citenamefont {Saffman}\ \emph {et~al.}(2010)\citenamefont
  {Saffman}, \citenamefont {Walker},\ and\ \citenamefont
  {M\o{}lmer}}]{saffman10a}%
  \BibitemOpen
  \bibfield  {author} {\bibinfo {author} {\bibfnamefont {M.}~\bibnamefont
  {Saffman}}, \bibinfo {author} {\bibfnamefont {T.~G.}\ \bibnamefont {Walker}},
  \ and\ \bibinfo {author} {\bibfnamefont {K.}~\bibnamefont {M\o{}lmer}},\
  }\href {\doibase 10.1103/RevModPhys.82.2313} {\bibfield  {journal} {\bibinfo
  {journal} {Rev. Mod. Phys.}\ }\textbf {\bibinfo {volume} {82}},\ \bibinfo
  {pages} {2313} (\bibinfo {year} {2010})}\BibitemShut {NoStop}%
\bibitem [{\citenamefont {Bluhm}\ \emph {et~al.}(1999)\citenamefont {Bluhm},
  \citenamefont {Kosteleck\'y},\ and\ \citenamefont {Russell}}]{bluhm99a}%
  \BibitemOpen
  \bibfield  {author} {\bibinfo {author} {\bibfnamefont {R.}~\bibnamefont
  {Bluhm}}, \bibinfo {author} {\bibfnamefont {V.~A.}\ \bibnamefont
  {Kosteleck\'y}}, \ and\ \bibinfo {author} {\bibfnamefont {N.}~\bibnamefont
  {Russell}},\ }\href {\doibase 10.1103/PhysRevLett.82.2254} {\bibfield
  {journal} {\bibinfo  {journal} {Phys. Rev. Lett.}\ }\textbf {\bibinfo
  {volume} {82}},\ \bibinfo {pages} {2254} (\bibinfo {year}
  {1999})}\BibitemShut {NoStop}%
\bibitem [{\citenamefont {Mills}\ and\ \citenamefont
  {Leventhal}(2002)}]{mills02a}%
  \BibitemOpen
  \bibfield  {author} {\bibinfo {author} {\bibfnamefont {A.~P.}\ \bibnamefont
  {Mills}, \bibfnamefont {Jr.}}\ and\ \bibinfo {author} {\bibfnamefont
  {M.}~\bibnamefont {Leventhal}},\ }\href {\doibase
  http://dx.doi.org/10.1016/S0168-583X(02)00789-9} {\bibfield  {journal}
  {\bibinfo  {journal} {Nucl. Instr. and Meth. in Phys. Res. B}\ }\textbf
  {\bibinfo {volume} {192}},\ \bibinfo {pages} {102 } (\bibinfo {year}
  {2002})}\BibitemShut {NoStop}%
\bibitem [{\citenamefont {Kellerbauer}\ \emph {et~al.}(2008)\citenamefont
  {Kellerbauer}, \citenamefont {Amoretti}, \citenamefont {Belov}, \citenamefont
  {Bonomi}, \citenamefont {Boscolo}, \citenamefont {Brusa}, \citenamefont
  {B{\"u}chner}, \citenamefont {Byakov}, \citenamefont {Cabaret}, \citenamefont
  {Canali}, \citenamefont {Carraro}, \citenamefont {Castelli}, \citenamefont
  {Cialdi}, \citenamefont {de~Combarieu}, \citenamefont {Comparat},
  \citenamefont {Consolati}, \citenamefont {Djourelov}, \citenamefont {Doser},
  \citenamefont {Drobychev}, \citenamefont {Dupasquier}, \citenamefont
  {Ferrari}, \citenamefont {Forget}, \citenamefont {Formaro}, \citenamefont
  {Gervasini}, \citenamefont {Giammarchi}, \citenamefont {Gninenko},
  \citenamefont {Gribakin}, \citenamefont {Hogan}, \citenamefont {Jacquey},
  \citenamefont {Lagomarsino}, \citenamefont {Manuzio}, \citenamefont
  {Mariazzi}, \citenamefont {Matveev}, \citenamefont {Meier}, \citenamefont
  {Merkt}, \citenamefont {Nedelec}, \citenamefont {Oberthaler}, \citenamefont
  {Pari}, \citenamefont {Prevedelli}, \citenamefont {Quasso}, \citenamefont
  {Rotondi}, \citenamefont {Sillou}, \citenamefont {Stepanov}, \citenamefont
  {Stroke}, \citenamefont {Testera}, \citenamefont {Tino}, \citenamefont
  {Tr{\'e}nec}, \citenamefont {Vairo}, \citenamefont {Vigu{\'e}}, \citenamefont
  {Walters}, \citenamefont {Warring}, \citenamefont {Zavatarelli},\ and\
  \citenamefont {Zvezhinskij}}]{kellerbauer08a}%
  \BibitemOpen
  \bibfield  {author} {\bibinfo {author} {\bibfnamefont {A.}~\bibnamefont
  {Kellerbauer}}, \bibinfo {author} {\bibfnamefont {M.}~\bibnamefont
  {Amoretti}}, \bibinfo {author} {\bibfnamefont {A.~S.}\ \bibnamefont {Belov}},
  \bibinfo {author} {\bibfnamefont {G.}~\bibnamefont {Bonomi}}, \bibinfo
  {author} {\bibfnamefont {I.}~\bibnamefont {Boscolo}}, \bibinfo {author}
  {\bibfnamefont {R.~S.}\ \bibnamefont {Brusa}}, \bibinfo {author}
  {\bibfnamefont {M.}~\bibnamefont {B{\"u}chner}}, \bibinfo {author}
  {\bibfnamefont {V.~M.}\ \bibnamefont {Byakov}}, \bibinfo {author}
  {\bibfnamefont {L.}~\bibnamefont {Cabaret}}, \bibinfo {author} {\bibfnamefont
  {C.}~\bibnamefont {Canali}}, \bibinfo {author} {\bibfnamefont
  {C.}~\bibnamefont {Carraro}}, \bibinfo {author} {\bibfnamefont
  {F.}~\bibnamefont {Castelli}}, \bibinfo {author} {\bibfnamefont
  {S.}~\bibnamefont {Cialdi}}, \bibinfo {author} {\bibfnamefont
  {M.}~\bibnamefont {de~Combarieu}}, \bibinfo {author} {\bibfnamefont
  {D.}~\bibnamefont {Comparat}}, \bibinfo {author} {\bibfnamefont
  {G.}~\bibnamefont {Consolati}}, \bibinfo {author} {\bibfnamefont
  {N.}~\bibnamefont {Djourelov}}, \bibinfo {author} {\bibfnamefont
  {M.}~\bibnamefont {Doser}}, \bibinfo {author} {\bibfnamefont
  {G.}~\bibnamefont {Drobychev}}, \bibinfo {author} {\bibfnamefont
  {A.}~\bibnamefont {Dupasquier}}, \bibinfo {author} {\bibfnamefont
  {G.}~\bibnamefont {Ferrari}}, \bibinfo {author} {\bibfnamefont
  {P.}~\bibnamefont {Forget}}, \bibinfo {author} {\bibfnamefont
  {L.}~\bibnamefont {Formaro}}, \bibinfo {author} {\bibfnamefont
  {A.}~\bibnamefont {Gervasini}}, \bibinfo {author} {\bibfnamefont {M.~G.}\
  \bibnamefont {Giammarchi}}, \bibinfo {author} {\bibfnamefont {S.~N.}\
  \bibnamefont {Gninenko}}, \bibinfo {author} {\bibfnamefont {G.}~\bibnamefont
  {Gribakin}}, \bibinfo {author} {\bibfnamefont {S.~D.}\ \bibnamefont {Hogan}},
  \bibinfo {author} {\bibfnamefont {M.}~\bibnamefont {Jacquey}}, \bibinfo
  {author} {\bibfnamefont {V.}~\bibnamefont {Lagomarsino}}, \bibinfo {author}
  {\bibfnamefont {G.}~\bibnamefont {Manuzio}}, \bibinfo {author} {\bibfnamefont
  {S.}~\bibnamefont {Mariazzi}}, \bibinfo {author} {\bibfnamefont {V.~A.}\
  \bibnamefont {Matveev}}, \bibinfo {author} {\bibfnamefont {J.~O.}\
  \bibnamefont {Meier}}, \bibinfo {author} {\bibfnamefont {F.}~\bibnamefont
  {Merkt}}, \bibinfo {author} {\bibfnamefont {P.}~\bibnamefont {Nedelec}},
  \bibinfo {author} {\bibfnamefont {M.~K.}\ \bibnamefont {Oberthaler}},
  \bibinfo {author} {\bibfnamefont {P.}~\bibnamefont {Pari}}, \bibinfo {author}
  {\bibfnamefont {M.}~\bibnamefont {Prevedelli}}, \bibinfo {author}
  {\bibfnamefont {F.}~\bibnamefont {Quasso}}, \bibinfo {author} {\bibfnamefont
  {A.}~\bibnamefont {Rotondi}}, \bibinfo {author} {\bibfnamefont
  {D.}~\bibnamefont {Sillou}}, \bibinfo {author} {\bibfnamefont {S.~V.}\
  \bibnamefont {Stepanov}}, \bibinfo {author} {\bibfnamefont {H.~H.}\
  \bibnamefont {Stroke}}, \bibinfo {author} {\bibfnamefont {G.}~\bibnamefont
  {Testera}}, \bibinfo {author} {\bibfnamefont {G.~M.}\ \bibnamefont {Tino}},
  \bibinfo {author} {\bibfnamefont {G.}~\bibnamefont {Tr{\'e}nec}}, \bibinfo
  {author} {\bibfnamefont {A.}~\bibnamefont {Vairo}}, \bibinfo {author}
  {\bibfnamefont {J.}~\bibnamefont {Vigu{\'e}}}, \bibinfo {author}
  {\bibfnamefont {H.}~\bibnamefont {Walters}}, \bibinfo {author} {\bibfnamefont
  {U.}~\bibnamefont {Warring}}, \bibinfo {author} {\bibfnamefont
  {S.}~\bibnamefont {Zavatarelli}}, \ and\ \bibinfo {author} {\bibfnamefont
  {D.~S.}\ \bibnamefont {Zvezhinskij}},\ }\href {\doibase
  http://dx.doi.org/10.1016/j.nimb.2007.12.010} {\bibfield  {journal} {\bibinfo
   {journal} {Nuc. Instr. and Meth. in Phys. Res. B}\ }\textbf {\bibinfo
  {volume} {266}},\ \bibinfo {pages} {351 } (\bibinfo {year}
  {2008})}\BibitemShut {NoStop}%
\bibitem [{\citenamefont {Cassidy}\ and\ \citenamefont
  {Hogan}(2014)}]{cassidy14a}%
  \BibitemOpen
  \bibfield  {author} {\bibinfo {author} {\bibfnamefont {D.~B.}\ \bibnamefont
  {Cassidy}}\ and\ \bibinfo {author} {\bibfnamefont {S.~D.}\ \bibnamefont
  {Hogan}},\ }\href {\doibase 10.1142/S2010194514602592} {\bibfield  {journal}
  {\bibinfo  {journal} {Int. J. of Mod. Phys.: Conf. Ser.}\ }\textbf {\bibinfo
  {volume} {30}},\ \bibinfo {pages} {1460259} (\bibinfo {year}
  {2014})}\BibitemShut {NoStop}%
\bibitem [{\citenamefont {Cassidy}\ \emph {et~al.}(2012)\citenamefont
  {Cassidy}, \citenamefont {Hisakado}, \citenamefont {Tom},\ and\ \citenamefont
  {Mills}}]{cassidy12a}%
  \BibitemOpen
  \bibfield  {author} {\bibinfo {author} {\bibfnamefont {D.~B.}\ \bibnamefont
  {Cassidy}}, \bibinfo {author} {\bibfnamefont {T.~H.}\ \bibnamefont
  {Hisakado}}, \bibinfo {author} {\bibfnamefont {H.~W.~K.}\ \bibnamefont
  {Tom}}, \ and\ \bibinfo {author} {\bibfnamefont {A.~P.}\ \bibnamefont
  {Mills}, \bibfnamefont {Jr.}},\ }\href {\doibase
  10.1103/PhysRevLett.108.043401} {\bibfield  {journal} {\bibinfo  {journal}
  {Phys. Rev. Lett.}\ }\textbf {\bibinfo {volume} {108}},\ \bibinfo {pages}
  {043401} (\bibinfo {year} {2012})}\BibitemShut {NoStop}%
\bibitem [{\citenamefont {Sprecher}\ \emph {et~al.}(2011)\citenamefont
  {Sprecher}, \citenamefont {Jungen}, \citenamefont {Ubachs},\ and\
  \citenamefont {Merkt}}]{sprecher11a}%
  \BibitemOpen
  \bibfield  {author} {\bibinfo {author} {\bibfnamefont {D.}~\bibnamefont
  {Sprecher}}, \bibinfo {author} {\bibfnamefont {C.}~\bibnamefont {Jungen}},
  \bibinfo {author} {\bibfnamefont {W.}~\bibnamefont {Ubachs}}, \ and\ \bibinfo
  {author} {\bibfnamefont {F.}~\bibnamefont {Merkt}},\ }\href {\doibase
  10.1039/c0fd00035c} {\bibfield  {journal} {\bibinfo  {journal} {Faraday
  Discuss.}\ }\textbf {\bibinfo {volume} {150}},\ \bibinfo {pages} {51}
  (\bibinfo {year} {2011})}\BibitemShut {NoStop}%
\bibitem [{\citenamefont {Hogan}(2013)}]{hogan13a}%
  \BibitemOpen
  \bibfield  {author} {\bibinfo {author} {\bibfnamefont {S.~D.}\ \bibnamefont
  {Hogan}},\ }\href {\doibase 10.1103/PhysRevA.87.063423} {\bibfield  {journal}
  {\bibinfo  {journal} {Phys. Rev. A}\ }\textbf {\bibinfo {volume} {87}},\
  \bibinfo {pages} {063423} (\bibinfo {year} {2013})}\BibitemShut {NoStop}%
\bibitem [{\citenamefont {H\"ansch}\ \emph {et~al.}(1975)\citenamefont
  {H\"ansch}, \citenamefont {Lee}, \citenamefont {Wallenstein},\ and\
  \citenamefont {Wieman}}]{Haensch75a}%
  \BibitemOpen
  \bibfield  {author} {\bibinfo {author} {\bibfnamefont {T.~W.}\ \bibnamefont
  {H\"ansch}}, \bibinfo {author} {\bibfnamefont {S.~A.}\ \bibnamefont {Lee}},
  \bibinfo {author} {\bibfnamefont {R.}~\bibnamefont {Wallenstein}}, \ and\
  \bibinfo {author} {\bibfnamefont {C.}~\bibnamefont {Wieman}},\ }\href
  {\doibase 10.1103/PhysRevLett.34.307} {\bibfield  {journal} {\bibinfo
  {journal} {Phys. Rev. Lett.}\ }\textbf {\bibinfo {volume} {34}},\ \bibinfo
  {pages} {307} (\bibinfo {year} {1975})}\BibitemShut {NoStop}%
\bibitem [{\citenamefont {Yamakita}\ \emph {et~al.}(2004)\citenamefont
  {Yamakita}, \citenamefont {Procter}, \citenamefont {Goodgame}, \citenamefont
  {Softley},\ and\ \citenamefont {Merkt}}]{yamakita04a}%
  \BibitemOpen
  \bibfield  {author} {\bibinfo {author} {\bibfnamefont {Y.}~\bibnamefont
  {Yamakita}}, \bibinfo {author} {\bibfnamefont {S.~R.}\ \bibnamefont
  {Procter}}, \bibinfo {author} {\bibfnamefont {A.~L.}\ \bibnamefont
  {Goodgame}}, \bibinfo {author} {\bibfnamefont {T.~P.}\ \bibnamefont
  {Softley}}, \ and\ \bibinfo {author} {\bibfnamefont {F.}~\bibnamefont
  {Merkt}},\ }\href {\doibase http://dx.doi.org/10.1063/1.1763146} {\bibfield
  {journal} {\bibinfo  {journal} {J. Chem. Phys.}\ }\textbf {\bibinfo {volume}
  {121}},\ \bibinfo {pages} {1419} (\bibinfo {year} {2004})}\BibitemShut
  {NoStop}%
\bibitem [{\citenamefont {Vliegen}\ \emph {et~al.}(2004)\citenamefont
  {Vliegen}, \citenamefont {W\"orner}, \citenamefont {Softley},\ and\
  \citenamefont {Merkt}}]{vliegen04a}%
  \BibitemOpen
  \bibfield  {author} {\bibinfo {author} {\bibfnamefont {E.}~\bibnamefont
  {Vliegen}}, \bibinfo {author} {\bibfnamefont {H.~J.}\ \bibnamefont
  {W\"orner}}, \bibinfo {author} {\bibfnamefont {T.~P.}\ \bibnamefont
  {Softley}}, \ and\ \bibinfo {author} {\bibfnamefont {F.}~\bibnamefont
  {Merkt}},\ }\href {\doibase 10.1103/PhysRevLett.92.033005} {\bibfield
  {journal} {\bibinfo  {journal} {Phys. Rev. Lett.}\ }\textbf {\bibinfo
  {volume} {92}},\ \bibinfo {pages} {033005} (\bibinfo {year}
  {2004})}\BibitemShut {NoStop}%
\bibitem [{\citenamefont {Hogan}\ and\ \citenamefont {Merkt}(2008)}]{hogan08a}%
  \BibitemOpen
  \bibfield  {author} {\bibinfo {author} {\bibfnamefont {S.~D.}\ \bibnamefont
  {Hogan}}\ and\ \bibinfo {author} {\bibfnamefont {F.}~\bibnamefont {Merkt}},\
  }\href {\doibase 10.1103/PhysRevLett.100.043001} {\bibfield  {journal}
  {\bibinfo  {journal} {Phys. Rev. Lett.}\ }\textbf {\bibinfo {volume} {100}},\
  \bibinfo {pages} {043001} (\bibinfo {year} {2008})}\BibitemShut {NoStop}%
\bibitem [{\citenamefont {Hogan}\ \emph {et~al.}(2009)\citenamefont {Hogan},
  \citenamefont {Seiler},\ and\ \citenamefont {Merkt}}]{hogan09a}%
  \BibitemOpen
  \bibfield  {author} {\bibinfo {author} {\bibfnamefont {S.~D.}\ \bibnamefont
  {Hogan}}, \bibinfo {author} {\bibfnamefont {{\relax Ch}.}~\bibnamefont
  {Seiler}}, \ and\ \bibinfo {author} {\bibfnamefont {F.}~\bibnamefont
  {Merkt}},\ }\href {\doibase 10.1103/PhysRevLett.103.123001} {\bibfield
  {journal} {\bibinfo  {journal} {Phys. Rev. Lett.}\ }\textbf {\bibinfo
  {volume} {103}},\ \bibinfo {pages} {123001} (\bibinfo {year}
  {2009})}\BibitemShut {NoStop}%
\bibitem [{\citenamefont {Seiler}\ \emph {et~al.}(2011)\citenamefont {Seiler},
  \citenamefont {Hogan},\ and\ \citenamefont {Merkt}}]{seiler11a}%
  \BibitemOpen
  \bibfield  {author} {\bibinfo {author} {\bibfnamefont {{\relax
  Ch}.}~\bibnamefont {Seiler}}, \bibinfo {author} {\bibfnamefont {S.~D.}\
  \bibnamefont {Hogan}}, \ and\ \bibinfo {author} {\bibfnamefont
  {F.}~\bibnamefont {Merkt}},\ }\href {\doibase 10.1039/C1CP21276A} {\bibfield
  {journal} {\bibinfo  {journal} {Phys. Chem. Chem. Phys.}\ }\textbf {\bibinfo
  {volume} {13}},\ \bibinfo {pages} {19000} (\bibinfo {year}
  {2011})}\BibitemShut {NoStop}%
\bibitem [{\citenamefont {Softley}(2004)}]{softley04a}%
  \BibitemOpen
  \bibfield  {author} {\bibinfo {author} {\bibfnamefont {T.~P.}\ \bibnamefont
  {Softley}},\ }\href {\doibase 10.1080/01442350310001652940} {\bibfield
  {journal} {\bibinfo  {journal} {Int. Rev. in Phys. Chem.}\ }\textbf {\bibinfo
  {volume} {23}},\ \bibinfo {pages} {1} (\bibinfo {year} {2004})}\BibitemShut
  {NoStop}%
\bibitem [{\citenamefont {Stoicheff}\ and\ \citenamefont
  {Weinberger}(1979)}]{stoicheff79a}%
  \BibitemOpen
  \bibfield  {author} {\bibinfo {author} {\bibfnamefont {B.}~\bibnamefont
  {Stoicheff}}\ and\ \bibinfo {author} {\bibfnamefont {E.}~\bibnamefont
  {Weinberger}},\ }\href@noop {} {\bibfield  {journal} {\bibinfo  {journal}
  {Can. J. Phys.}\ }\textbf {\bibinfo {volume} {57}},\ \bibinfo {pages} {2143}
  (\bibinfo {year} {1979})}\BibitemShut {NoStop}%
\bibitem [{\citenamefont {Penent}\ \emph {et~al.}(1988)\citenamefont {Penent},
  \citenamefont {Delande},\ and\ \citenamefont {Gay}}]{penent88a}%
  \BibitemOpen
  \bibfield  {author} {\bibinfo {author} {\bibfnamefont {F.}~\bibnamefont
  {Penent}}, \bibinfo {author} {\bibfnamefont {D.}~\bibnamefont {Delande}}, \
  and\ \bibinfo {author} {\bibfnamefont {J.~C.}\ \bibnamefont {Gay}},\ }\href
  {\doibase 10.1103/PhysRevA.37.4707} {\bibfield  {journal} {\bibinfo
  {journal} {Phys. Rev. A}\ }\textbf {\bibinfo {volume} {37}},\ \bibinfo
  {pages} {4707} (\bibinfo {year} {1988})}\BibitemShut {NoStop}%
\bibitem [{\citenamefont {G\'en\'evriez}\ \emph {et~al.}(2014)\citenamefont
  {G\'en\'evriez}, \citenamefont {Urbain}, \citenamefont {Brouri},
  \citenamefont {O'Connor}, \citenamefont {Dunseath},\ and\ \citenamefont
  {Terao-Dunseath}}]{genevriez14a}%
  \BibitemOpen
  \bibfield  {author} {\bibinfo {author} {\bibfnamefont {M.}~\bibnamefont
  {G\'en\'evriez}}, \bibinfo {author} {\bibfnamefont {X.}~\bibnamefont
  {Urbain}}, \bibinfo {author} {\bibfnamefont {M.}~\bibnamefont {Brouri}},
  \bibinfo {author} {\bibfnamefont {A.~P.}\ \bibnamefont {O'Connor}}, \bibinfo
  {author} {\bibfnamefont {K.~M.}\ \bibnamefont {Dunseath}}, \ and\ \bibinfo
  {author} {\bibfnamefont {M.}~\bibnamefont {Terao-Dunseath}},\ }\href
  {\doibase 10.1103/PhysRevA.89.053430} {\bibfield  {journal} {\bibinfo
  {journal} {Phys. Rev. A}\ }\textbf {\bibinfo {volume} {89}},\ \bibinfo
  {pages} {053430} (\bibinfo {year} {2014})}\BibitemShut {NoStop}%
\bibitem [{\citenamefont {Halfmann}\ \emph {et~al.}(2000)\citenamefont
  {Halfmann}, \citenamefont {Koensgen},\ and\ \citenamefont
  {Bergmann}}]{Halfmann00}%
  \BibitemOpen
  \bibfield  {author} {\bibinfo {author} {\bibfnamefont {T.}~\bibnamefont
  {Halfmann}}, \bibinfo {author} {\bibfnamefont {J.}~\bibnamefont {Koensgen}},
  \ and\ \bibinfo {author} {\bibfnamefont {K.}~\bibnamefont {Bergmann}},\
  }\href {http://stacks.iop.org/0957-0233/11/i=10/a=312} {\bibfield  {journal}
  {\bibinfo  {journal} {Meas. Sci. and Tech.}\ }\textbf {\bibinfo {volume}
  {11}},\ \bibinfo {pages} {1510} (\bibinfo {year} {2000})}\BibitemShut
  {NoStop}%
\bibitem [{\citenamefont {Deissler}\ \emph {et~al.}(2008)\citenamefont
  {Deissler}, \citenamefont {Hughes}, \citenamefont {Burke},\ and\
  \citenamefont {Sackett}}]{Deissler08a}%
  \BibitemOpen
  \bibfield  {author} {\bibinfo {author} {\bibfnamefont {B.}~\bibnamefont
  {Deissler}}, \bibinfo {author} {\bibfnamefont {K.~J.}\ \bibnamefont
  {Hughes}}, \bibinfo {author} {\bibfnamefont {J.~H.~T.}\ \bibnamefont
  {Burke}}, \ and\ \bibinfo {author} {\bibfnamefont {C.~A.}\ \bibnamefont
  {Sackett}},\ }\href {\doibase 10.1103/PhysRevA.77.031604} {\bibfield
  {journal} {\bibinfo  {journal} {Phys. Rev. A}\ }\textbf {\bibinfo {volume}
  {77}},\ \bibinfo {pages} {031604} (\bibinfo {year} {2008})}\BibitemShut
  {NoStop}%
\bibitem [{\citenamefont {Zare}(1988)}]{zarebook}%
  \BibitemOpen
  \bibfield  {author} {\bibinfo {author} {\bibfnamefont {R.~N.}\ \bibnamefont
  {Zare}},\ }\href@noop {} {\emph {\bibinfo {title} {Angular Momentum:
  Understanding Spatial Aspects in Chemistry and Physics}}}\ (\bibinfo
  {publisher} {Wiley-Blackwell},\ \bibinfo {year} {1988})\BibitemShut {NoStop}%
\bibitem [{\citenamefont {Zimmerman}\ \emph {et~al.}(1979)\citenamefont
  {Zimmerman}, \citenamefont {Littman}, \citenamefont {Kash},\ and\
  \citenamefont {Kleppner}}]{zimmerman79a}%
  \BibitemOpen
  \bibfield  {author} {\bibinfo {author} {\bibfnamefont {M.~L.}\ \bibnamefont
  {Zimmerman}}, \bibinfo {author} {\bibfnamefont {M.~G.}\ \bibnamefont
  {Littman}}, \bibinfo {author} {\bibfnamefont {M.~M.}\ \bibnamefont {Kash}}, \
  and\ \bibinfo {author} {\bibfnamefont {D.}~\bibnamefont {Kleppner}},\ }\href
  {\doibase 10.1103/PhysRevA.20.2251} {\bibfield  {journal} {\bibinfo
  {journal} {Phys. Rev. A}\ }\textbf {\bibinfo {volume} {20}},\ \bibinfo
  {pages} {2251} (\bibinfo {year} {1979})}\BibitemShut {NoStop}%
\bibitem [{\citenamefont {Martin}(1987)}]{martin87a}%
  \BibitemOpen
  \bibfield  {author} {\bibinfo {author} {\bibfnamefont {W.~C.}\ \bibnamefont
  {Martin}},\ }\href {\doibase 10.1103/PhysRevA.36.3575} {\bibfield  {journal}
  {\bibinfo  {journal} {Phys. Rev. A}\ }\textbf {\bibinfo {volume} {36}},\
  \bibinfo {pages} {3575} (\bibinfo {year} {1987})}\BibitemShut {NoStop}%
\bibitem [{\citenamefont {Farley}\ and\ \citenamefont
  {Wing}(1981)}]{farley81a}%
  \BibitemOpen
  \bibfield  {author} {\bibinfo {author} {\bibfnamefont {J.~W.}\ \bibnamefont
  {Farley}}\ and\ \bibinfo {author} {\bibfnamefont {W.~H.}\ \bibnamefont
  {Wing}},\ }\href {\doibase 10.1103/PhysRevA.23.2397} {\bibfield  {journal}
  {\bibinfo  {journal} {Phys. Rev. A}\ }\textbf {\bibinfo {volume} {23}},\
  \bibinfo {pages} {2397} (\bibinfo {year} {1981})}\BibitemShut {NoStop}%
\bibitem [{\citenamefont {Mills}(1979)}]{Mills79}%
  \BibitemOpen
  \bibfield  {author} {\bibinfo {author} {\bibfnamefont {A.~P.}\ \bibnamefont
  {Mills}, \bibfnamefont {Jr}},\ }\href@noop {} {\bibfield  {journal} {\bibinfo
   {journal} {Solid State Comm.}\ }\textbf {\bibinfo {volume} {31}},\ \bibinfo
  {pages} {623} (\bibinfo {year} {1979})}\BibitemShut {NoStop}%
\bibitem [{\citenamefont {Cassidy}\ \emph {et~al.}(2010)\citenamefont
  {Cassidy}, \citenamefont {Crivelli}, \citenamefont {Hisakado}, \citenamefont
  {Liszkay}, \citenamefont {Meligne}, \citenamefont {Perez}, \citenamefont
  {Tom},\ and\ \citenamefont {Mills}}]{cassidy2010}%
  \BibitemOpen
  \bibfield  {author} {\bibinfo {author} {\bibfnamefont {D.~B.}\ \bibnamefont
  {Cassidy}}, \bibinfo {author} {\bibfnamefont {P.}~\bibnamefont {Crivelli}},
  \bibinfo {author} {\bibfnamefont {T.~H.}\ \bibnamefont {Hisakado}}, \bibinfo
  {author} {\bibfnamefont {L.}~\bibnamefont {Liszkay}}, \bibinfo {author}
  {\bibfnamefont {V.~E.}\ \bibnamefont {Meligne}}, \bibinfo {author}
  {\bibfnamefont {P.}~\bibnamefont {Perez}}, \bibinfo {author} {\bibfnamefont
  {H.~W.~K.}\ \bibnamefont {Tom}}, \ and\ \bibinfo {author} {\bibfnamefont
  {A.~P.}\ \bibnamefont {Mills}, \bibfnamefont {Jr.}},\ }\href {\doibase
  10.1103/PhysRevA.81.012715} {\bibfield  {journal} {\bibinfo  {journal} {Phys.
  Rev. A}\ }\textbf {\bibinfo {volume} {81}},\ \bibinfo {pages} {012715}
  (\bibinfo {year} {2010})}\BibitemShut {NoStop}%
\bibitem [{\citenamefont {Jones}\ \emph {et~al.}(2014)\citenamefont {Jones},
  \citenamefont {Hisakado}, \citenamefont {Goldman}, \citenamefont {Tom},
  \citenamefont {Mills},\ and\ \citenamefont {Cassidy}}]{Jones14}%
  \BibitemOpen
  \bibfield  {author} {\bibinfo {author} {\bibfnamefont {A.~C.~L.}\
  \bibnamefont {Jones}}, \bibinfo {author} {\bibfnamefont {T.~H.}\ \bibnamefont
  {Hisakado}}, \bibinfo {author} {\bibfnamefont {H.~J.}\ \bibnamefont
  {Goldman}}, \bibinfo {author} {\bibfnamefont {H.~W.~K.}\ \bibnamefont {Tom}},
  \bibinfo {author} {\bibfnamefont {A.~P.}\ \bibnamefont {Mills}, \bibfnamefont
  {Jr.}}, \ and\ \bibinfo {author} {\bibfnamefont {D.~B.}\ \bibnamefont
  {Cassidy}},\ }\href {\doibase 10.1103/PhysRevA.90.012503} {\bibfield
  {journal} {\bibinfo  {journal} {Phys. Rev. A}\ }\textbf {\bibinfo {volume}
  {90}},\ \bibinfo {pages} {012503} (\bibinfo {year} {2014})}\BibitemShut
  {NoStop}%
\bibitem [{\citenamefont {Fee}\ \emph {et~al.}(1993)\citenamefont {Fee},
  \citenamefont {Chu}, \citenamefont {Mills}, \citenamefont {Chichester},
  \citenamefont {Zuckerman}, \citenamefont {Shaw},\ and\ \citenamefont
  {Danzmann}}]{Fee93}%
  \BibitemOpen
  \bibfield  {author} {\bibinfo {author} {\bibfnamefont {M.~S.}\ \bibnamefont
  {Fee}}, \bibinfo {author} {\bibfnamefont {S.}~\bibnamefont {Chu}}, \bibinfo
  {author} {\bibfnamefont {A.~P.}\ \bibnamefont {Mills}, \bibfnamefont {Jr.}},
  \bibinfo {author} {\bibfnamefont {R.~J.}\ \bibnamefont {Chichester}},
  \bibinfo {author} {\bibfnamefont {D.~M.}\ \bibnamefont {Zuckerman}}, \bibinfo
  {author} {\bibfnamefont {E.~D.}\ \bibnamefont {Shaw}}, \ and\ \bibinfo
  {author} {\bibfnamefont {K.}~\bibnamefont {Danzmann}},\ }\href {\doibase
  10.1103/PhysRevA.48.192} {\bibfield  {journal} {\bibinfo  {journal} {Phys.
  Rev. A}\ }\textbf {\bibinfo {volume} {48}},\ \bibinfo {pages} {192} (\bibinfo
  {year} {1993})}\BibitemShut {NoStop}%
\bibitem [{\citenamefont {Vliegen}\ and\ \citenamefont
  {Merkt}(2006)}]{vliegen2006}%
  \BibitemOpen
  \bibfield  {author} {\bibinfo {author} {\bibfnamefont {E.}~\bibnamefont
  {Vliegen}}\ and\ \bibinfo {author} {\bibfnamefont {F.}~\bibnamefont
  {Merkt}},\ }\href {\doibase 10.1103/PhysRevLett.97.033002} {\bibfield
  {journal} {\bibinfo  {journal} {Phys. Rev. Lett.}\ }\textbf {\bibinfo
  {volume} {97}},\ \bibinfo {pages} {033002} (\bibinfo {year}
  {2006})}\BibitemShut {NoStop}%
\bibitem [{\citenamefont {Hogan}\ \emph {et~al.}(2011)\citenamefont {Hogan},
  \citenamefont {Motsch},\ and\ \citenamefont {Merkt}}]{hogan11a}%
  \BibitemOpen
  \bibfield  {author} {\bibinfo {author} {\bibfnamefont {S.~D.}\ \bibnamefont
  {Hogan}}, \bibinfo {author} {\bibfnamefont {M.}~\bibnamefont {Motsch}}, \
  and\ \bibinfo {author} {\bibfnamefont {F.}~\bibnamefont {Merkt}},\ }\href
  {\doibase 10.1039/C1CP21733J} {\bibfield  {journal} {\bibinfo  {journal}
  {Phys. Chem. Chem. Phys.}\ }\textbf {\bibinfo {volume} {13}},\ \bibinfo
  {pages} {18705} (\bibinfo {year} {2011})}\BibitemShut {NoStop}%
\bibitem [{\citenamefont {Haas}\ \emph {et~al.}(2006)\citenamefont {Haas},
  \citenamefont {Jentschura}, \citenamefont {Keitel}, \citenamefont
  {Kolachevsky}, \citenamefont {Herrmann}, \citenamefont {Fendel},
  \citenamefont {Fischer}, \citenamefont {Udem}, \citenamefont {Holzwarth},
  \citenamefont {H\"ansch}, \citenamefont {Scully},\ and\ \citenamefont
  {Agarwal}}]{Haas2006}%
  \BibitemOpen
  \bibfield  {author} {\bibinfo {author} {\bibfnamefont {M.}~\bibnamefont
  {Haas}}, \bibinfo {author} {\bibfnamefont {U.~D.}\ \bibnamefont
  {Jentschura}}, \bibinfo {author} {\bibfnamefont {C.~H.}\ \bibnamefont
  {Keitel}}, \bibinfo {author} {\bibfnamefont {N.}~\bibnamefont {Kolachevsky}},
  \bibinfo {author} {\bibfnamefont {M.}~\bibnamefont {Herrmann}}, \bibinfo
  {author} {\bibfnamefont {P.}~\bibnamefont {Fendel}}, \bibinfo {author}
  {\bibfnamefont {M.}~\bibnamefont {Fischer}}, \bibinfo {author} {\bibfnamefont
  {{\relax Th}.}~\bibnamefont {Udem}}, \bibinfo {author} {\bibfnamefont
  {R.}~\bibnamefont {Holzwarth}}, \bibinfo {author} {\bibfnamefont {T.~W.}\
  \bibnamefont {H\"ansch}}, \bibinfo {author} {\bibfnamefont {M.~O.}\
  \bibnamefont {Scully}}, \ and\ \bibinfo {author} {\bibfnamefont {G.~S.}\
  \bibnamefont {Agarwal}},\ }\href {\doibase 10.1103/PhysRevA.73.052501}
  {\bibfield  {journal} {\bibinfo  {journal} {Phys. Rev. A}\ }\textbf {\bibinfo
  {volume} {73}},\ \bibinfo {pages} {052501} (\bibinfo {year}
  {2006})}\BibitemShut {NoStop}%
\bibitem [{\citenamefont {Cassidy}\ \emph {et~al.}(2006)\citenamefont
  {Cassidy}, \citenamefont {Deng}, \citenamefont {Greaves},\ and\ \citenamefont
  {Mills}}]{Cassidy2006}%
  \BibitemOpen
  \bibfield  {author} {\bibinfo {author} {\bibfnamefont {D.~B.}\ \bibnamefont
  {Cassidy}}, \bibinfo {author} {\bibfnamefont {S.~H.~M.}\ \bibnamefont
  {Deng}}, \bibinfo {author} {\bibfnamefont {R.~G.}\ \bibnamefont {Greaves}}, \
  and\ \bibinfo {author} {\bibfnamefont {A.~P.}\ \bibnamefont {Mills},
  \bibfnamefont {Jr.}},\ }\href {\doibase 10.1063/1.2221509} {\bibfield
  {journal} {\bibinfo  {journal} {Rev. Sci. Instrum.}\ }\textbf {\bibinfo
  {volume} {77}},\ \bibinfo {pages} {073106} (\bibinfo {year}
  {2006})}\BibitemShut {NoStop}%
\bibitem [{\citenamefont {Mills}(1980)}]{Mills1980}%
  \BibitemOpen
  \bibfield  {author} {\bibinfo {author} {\bibfnamefont {A.~P.}\ \bibnamefont
  {Mills}, \bibfnamefont {Jr.}},\ }\href {\doibase 10.1007/BF00899716}
  {\bibfield  {journal} {\bibinfo  {journal} {Appl. Phys.}\ }\textbf {\bibinfo
  {volume} {23}},\ \bibinfo {pages} {189} (\bibinfo {year} {1980})}\BibitemShut
  {NoStop}%
\end{thebibliography}%

\appendix*
\section{Lineshape simulation}\label{simAppendix}
The calculation of the lineshapes associated with the spectral features observed in the experiments reported here was particularly important in understanding the role of ac Stark shifts, and the effects of the properties of the laser and atomic beams in the experiments. The features in the spectra presented are broadened as a result of ac Stark shifts of the $2\,^3\text{S}_{1}$ state, leading to asymmetric lineshapes with widths greater than those expected from the bandwidth of the laser. The precise shape of these features was determined by the overlap of the spatial distribution of atoms in the supersonic beam with the intensity distribution of the focussed laser beam. 

The laser beam was focussed such that where it intersected the atomic beam axis its intensity profile was Gaussian with a width $\sigma_y(x=0)\simeq\sigma_z(x=0)=70~\mu$m, where $x$ is the direction of propagation of the laser beam, and $z$ is the direction of propagation of the atomic beam (see Fig.~\ref{fig:expSetup}). Because of the limited precision, $\pm1$~mm, with which the crossing point of the atomic beam and the laser beam in the $x$-dimension could be determined in the experiments, the error on the experimental value for $\sigma_y(x=0)$ and $\sigma_z(x=0)$ was $\pm11~\mu$m. The numerical value of $70~\mu$m used in the simulation of the lineshapes represented the best fit to the experimental data within the range of values encompassed by these error bars.

The spatial distribution of atoms in the supersonic beam was defined by the geometry of the pulsed valve and the 2~mm-diameter skimmer. At the position where the atomic beam intersected the laser beam it occupied a significantly greater volume than the laser beam. Traveling with a mean forward speed of $v_z=1950~$ms$^{-1}$, the distance the atoms travelled during the time that the laser beam was present was negligible, and it was therefore assumed that the atoms were stationary during excitation. The calculations presented here involved the generation of ensembles of atoms with uniform distributions in the $y$ and $z$ dimensions, perpendicular to the direction of propagation of the laser beam, and a Gaussian distribution (with a width $\sigma_x=1.5~$mm) in the $x$-direction, along which the laser beam propagated. The width of this Gaussian distribution was determined from the geometry of the experimental apparatus, in particular the relative positions of the pulsed valve, the skimmer and the photoexcitation region. As the transverse temperature of the atomic beam was not precisely known, the numerical value of $\sigma_x$ employed in the lineshape simulations was adjusted within a range of 15\% of the value determined geometrically to realize the best fit of the calculated lineshapes with the those measured in the experiments. In the $y$- and $z$-directions the spatial extent of the ensemble of atoms was sufficiently large when compared to the size of the converging laser beam that the assumption of a uniform distribution of atoms was reasonable. Including the correct distribution of atoms in the $x$-direction was crucial for the accurate calculation of the experimental line shapes.

\begin{figure}[t]
\centerline{\includegraphics[width=1 \linewidth]{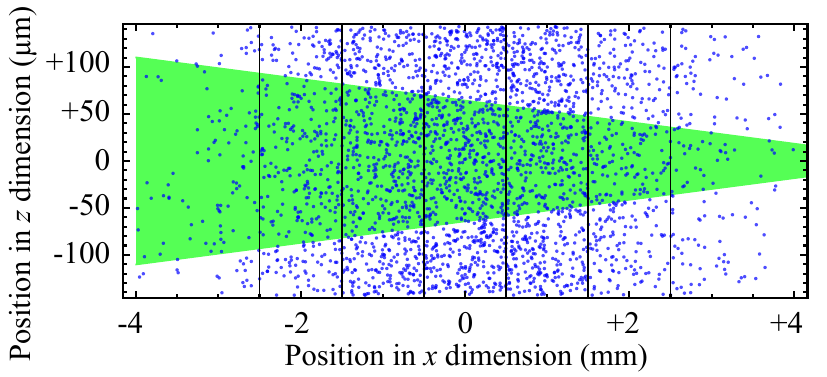}}
\caption{(Color online) A schematic diagram showing the overlap of the atomic cloud (blue points) with the converging laser beam (green shaded region). The atomic distribution in the $xz$-plane, where the distribution has a Gaussian shape along $x$ with width $\sigma_x=1.5~$mm and centroid $x_0=0~$mm. The black lines show the positions of the $yz-$planes described in the text.}
\label{fig:appFigDist}
\end{figure}

A three-dimensional laser intensity distribution was generated by making detailed beam profile and energy measurements. Every atom in the calculations was assigned a laser intensity depending on its position, $I_{\text{laser}}(x,y,z)$ from which the ac Stark shift was then calculated. This shift was then convolved with the laser bandwidth, a Gaussian function with width $\sigma=5.0\sqrt{2}~$GHz (where the factor of $\sqrt{2}$ accounts for the two-photon nature of the transitions). The two-photon transition strength for each atom was then included by multiplying the amplitude of each convolved single atom spectrum with $I_{\text{laser}}(x,y,z)^2$. The validity of this approach is based upon the $I_{\text{laser}}(x,y,z)^2$ dependence of the integrated signal associated with the transition to each individual Rydberg state, an example of which is presented in Fig.~\ref{fig:intVSenergy}. The lineshapes for each atom were then summed to generate the overall lineshape. The calculated lineshapes are in excellent agreement with those in the experimental spectra, demonstrating that the line profiles resulting from the ac Stark shift of the $2\,^3\text{S}_{1}$ state are therefore, under the conditions of the experiment, independent of $n$ (see Fig.~\ref{fig:peakWidthsN}), while the exact shape of the spectral features was determined by the spatial overlap of the atomic and laser beams.

\begin{figure}[t]
\centerline{\includegraphics[width=1 \linewidth]{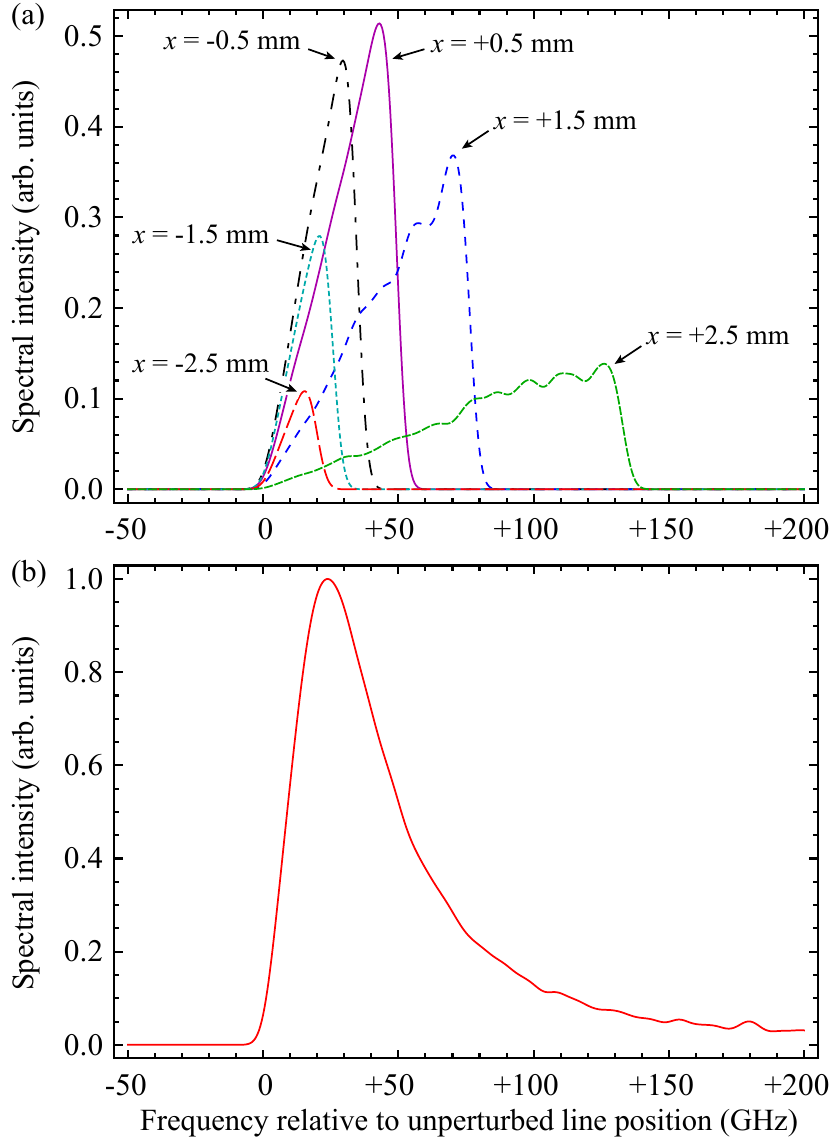}}
\caption{(Color online) (a) Line profiles calculated for atoms located in $yz$-planes with positions in the $x$-dimension as indicated. These planes are shown as black lines in Fig \ref{fig:appFigDist}. (b) The line profiles generated by summing the spectral contributions from a continuous three-dimensional atomic ensemble.}
\label{fig:appFigPlanes}
\end{figure}

One surprising feature of the recorded lineshapes is the strong contribution from atoms that experience only a weak ac Stark shift. This is particularly striking in the data in Fig.~\ref{fig:sigVSenergy}, where, as the laser energy is increased the features broaden as the ac Stark shifts increase, while the position of the maximum of each feature remains close to the unperturbed line position, and the signal from weakly-shifted atoms is observed to increase with laser intensity. This spectral behaviour is explained by considering the contributions to the overall lineshape from atoms at different positions in the $x$-dimension. In Fig.~\ref{fig:appFigDist} a schematic diagram highlighting the overlap of the atoms with the laser beam in the $xz-$plane is presented. The green shaded region represents the area enclosed by the $2\sigma$ full-width of the laser beam as it converges in the $x$-direction, and the blue dots represent the cloud of metastable He atoms in the 2\,$^3\text{S}_1$ state at the instant that the laser beam is present. The spatial contribution to the spectral profiles are seen by calculating the $29$d lineshape associated with atoms in a number of $yz$-planes located at selected positions in the $x$-dimension (black vertical lines in Fig.~\ref{fig:appFigDist}). Fig.~\ref{fig:appFigPlanes}(a) shows the lineshapes calculated for atoms in planes located at $x=+2.5$, $+1.5$, $+0.5$, $-0.5$, $-1.5$ and $-2.5~$mm. Each plane contributes an asymmetric lineshape, with a maximum at a frequency greater than the unperturbed line position, close to the transition frequency corresponding to the greatest ac Stark shift in that plane. This shape stems from the fact that in a given plane the greatest signal always comes from those atoms near the centre of the laser beam, where the excitation probability is highest. These atoms also experience the greatest ac Stark shifts, thus generating the asymmetric lineshapes, shifted to higher frequency. This effect is seen most clearly in Fig.~\ref{fig:appFigPlanes}(a) in the line profile from the atoms in the $x=+2.5~$mm plane, which is located closest to the laser focus, where the peak intensity $I_{\text{laser}}(2.5~\text{mm},0,0)=1.5\times10^{14}~$Wm$^{-2}$ (resulting in a maximal ac Stark shift of 133~GHz). This lineshape is very asymmetric, with a maximum $\sim$130~GHz above the unperturbed line position. In planes further from the laser focus the peak intensities are lower, and the lineshapes shift towards the unperturbed position, becoming more symmetric as they do so. Since the peak intensity varies quadratically in the $x$-dimension the contributions from the planes located further from the laser focus bunch up and overlap close to the unperturbed line position, creating a large spectral peak near this frequency, while the atoms that are positioned closer to the laser focus experience a wide range of intensities, creating a broad spectral signal spread over a wide range of frequencies. This is clearly seen in the calculated spectral profile shown in Fig.~\ref{fig:appFigPlanes}(b), which was calculated with a continuous three-dimensional atomic ensemble.

\end{document}